\pdfoutput=1
\documentclass[12pt,eqsecnum,preprint]{aastex}
\title{Three Dimensional Simulations of Vertical Magnetic Flux in the Immediate Vicinity of Black Holes}
\begin{document}
\author{Brian Punsly\altaffilmark{1}, Igor V. Igumenshchev \altaffilmark{2}, Shigenobu Hirose\altaffilmark{3}}
\altaffiltext{1}{4014 Emerald Street No.116, Torrance CA, USA 90503
and International Center for Relativistic Astrophysics, ICRANet,
Piazza della Repubblica 10 Pescara 65100, Italy}
\altaffiltext{2}{Laboratory for Laser Energetics, University of Rochester\\
250 East River Road, Rochester, NY 14623} \altaffiltext{3}{The Earth
Simulator Center, JAMSTEC 3173-25 Showamachi, Kanazawa-ku, Yokohama,
Kanagawa 236-0001, Japan}
\email{brian.punsly@verizon.net}
\begin{abstract}
This article reports on three-dimensional (3-D) MHD simulations of
non-rotating and rapidly rotating black holes and the adjacent black
hole accretion disk magnetospheres. A particular emphasis is placed
on the vertical magnetic flux that is advected inward from large
radii and threads the equatorial plane near the event horizon. In
both cases of non-rotating and rotating black holes, the existence
of a significant vertical magnetic field in this region is like a
switch that creates powerful jets. There are many similarities in
the vertical flux dynamics in these two cases in spite of the
tremendous enhancement of azimuthal twisting of the field lines and
enhancement of the jet power because of an ``ergospheric disk" in
the Kerr metric. A 3-D approach is essential because two-dimensional
axisymmetric flows are incapable of revealing the nature of vertical
flux near a black hole. Poloidal field lines from the ergospheric
accretion region have been visualized in 3-D and much of the article
is devoted to a formal classification of the different
manifestations of vertical flux in the Kerr case.
\end{abstract}

\section{Introduction}It is widely believed that magnetized plasma ejected from the
vicinity of a supermassive black hole is the origin of the
relativistic jets that have been detected in extragalactic radio
sources. Furthermore, such jets are believed to be highly magnetized
at their launch point in order for jet plasma to achieve highly
relativistic velocities. One of the biggest controversies in the
theory of relativistic, extragalactic jets is whether they initiate
from local turbulent magnetic fields created in an accretion flow
or from a large scale reservoir of
poloidal flux that is advected inward from large distances. The
implication of the first scenario is that the coherent field within
the dense gas is inhibited by the turbulent magnetic diffusivity of
the plasma \citep{van89,lub94}. As a corollary to this, it
customarily concluded that the accretion disk cannot maintain a
strong magnetosphere and therefore the jet power would be quite
small relative to FRII (Fanaroff-Riley type II) radio sources
\citep{pra97}.\footnote{these arguments have been challenged by
\citet{rey06}} The only place in which a strong field can exist in
this scenario is pinned against the event horizon, in the vortex of
the accretion flow, and then only under a select set of
circumstances \citep{bec08}. Conversely, the second scenario implies
that large scale poloidal magnetic fields can be maintained by the
accreting gas all the way down to the black hole
\citep{bis74,bis76,nar03}. Provided that the turbulence in the
accretion flow does not induce an outward diffusion of magnetic flux
that is too rapid, the inner regions of the accretion disk will
support significant vertical flux \citep{rot08}. In the weak field
limit, modest mildly relativistic jets can be driven by
magneto-centrifugal forces \citep{bla82}. If a strong vertical field
is accumulated in the inner regions of the disk, a relativistic
Poynting jet forms that is far more powerful than its
magneto-centrifugal counterpart \citep{lov76,bla76,ust00}. If the
vertical flux is accumulated within the plunge region, an
ergospheric disk and the associated Poynting jet can form
\citep{pun90,pun01}. This is of particular interest to the study of
powerful jets in radio loud AGN since the jet power is increased an
order of magnitude over the Poynting jets from flux that is merely
pinned to the event horizon of a rapidly rotating black hole.
\par The longstanding controversy over the existence of significant
vertical flux in black hole accretion disks is rooted in a legacy of
theoretical studies. The time evolution of vertical flux depends on
critical assumptions about the accreting gas that are not well
understood in a quasar environment
\citep{mei04,spr05,shi90,kat04,rot08}. Numerical methods have
the tremendous advantage of being able to depict the very nonlinear
magnetic flux evolution of scenarios based on said assumptions -
something that can only be addressed with crude approximations
theoretically. Given that the potential existence of vertical flux
in the equatorial accretion flow is fundamental to the theory of
black hole driven jets, we use the data from long term 3-D
simulations, that contain significant vertical flux near the black
hole, in order to shed light on the general properties of these flux
distributions. This paper is not an overview of the set of numerical
assumptions, initial conditions and black hole states that give rise
to such features. This is far beyond the state of this burgeoning
field of numerical work. Instead it reports on the first two known
anecdotal cases of long term 3-D simulations that show significant
magnetic flux adjacent to the black hole. As a pioneering work in
this area, it is by no means complete. Our particular emphasis is on
exploring the possible varieties of self-consistent MHD structures
that can support significant vertical flux near black holes. For our
purposes, these simulations offer a wealth of information and
revelations.
\par This paper
employs both 2-D and 3-D numerical simulations to investigate the
nature of vertical flux near black holes. We explore two extreme
versions of a black hole. First, a non-rotating black hole in the
Pseudo-Newtonian approximation with a continuous net magnetic flux
accretion (section 3). Additionally, we look at a near maximally
rotating black hole (a/M = 0.99, where "a" is the angular momentum
per unit mass, "M", of the black hole in geometrized units) with
zero net flux in the initial state and no flux accretion from the
outer boundary afterwards (section 4). Comparing and contrasting
these simulations in section 5 offers new insights into the nature
of vertical flux near black holes.
\par Perfect magnetohydrodynamic (simply MHD
hereafter) numerical simulations offer us a virtual laboratory for
exploring the possible self-consistent MHD structures that
might exist near black holes. We consider this tool only as such and
we do not offer them any more import than this. Thus, it is
necessary to stress that we do not claim
\begin{enumerate}

\item that the simulations that are used to determine these results are representative of actual AGN central
engines, nor

\item that the results that we see in the simulation are independent of the initial conditions and the assumptions of the numerical
code, nor

\item that the perfect MHD assumption implemented in the numerical code is applicable
to the extremely hot and very turbulent plasma in the inner
accretion flow. In particular, the manner in which flux diffuses and
reconnects, nor

\item that the results of these simulations are not influenced by numerical
diffusion.

\end{enumerate}
\par A second goal of this study is to consider visualization
techniques for studying complicated magnetospheric phenomena. For
example, it has been discovered that the base of the solar
magnetosphere is filled with powerful, complicated geometrical
structures such as bundles of field lines that form bent
semi-circular arcs in which the individual lines twist around each
other. As with the sun, the complexity of the 3-D structures can not
be readily understood with 2-D visualization techniques
\citep{dor99,arc08,war08}. We anticipate that the inner parts of an
accretion flow forming a quasar should have a magnetosphere with
similar or greater complexity since it is extremely turbulent and
hotter than the solar chromosphere (there is also strong azimuthal
shearing due to the differentially rotating accretion flow that is
far more severe than in the solar atmosphere). We have
developed certain graphical techniques within a 3-D visualization
platform, Paraview 3.3.0. We
show that there is a tremendous distinction between azimuthally
averaged plots and the 3-D visualization. The actual field line
geometry is virtually impossible to figure out from the 2-D averages
because the magnetosphere is very inhomogeneous and twisted. We
found it necessary to use 3-D animations of a camera revolving
around the black hole in order to get a truly clear picture since
there are so many distortions from perspective and there are always
lines of sight that are blocked by other field lines, obscuring gas
or the black hole itself. In section 4, we use our 3-D visualization
techniques to describe four distinct vertical field line varieties
that arise in simulations of the innermost accretion flow near a
rapidly rotating black hole.
\section{A Fundamental Topological Issue}In this section, we comment
on how different initial conditions and boundary conditions can
suppress or enhance, the amount of vertical flux near a black hole.
We believe that the presence of vertical flux in the inner accretion
flow is rendered to a basic topological consideration that we
motivate in this discussion and figure 1.
\par The simulations in \citet{igu08} considered a net vertical flux that
accretes towards a black hole in both 2-D and 3-D from the outer
regions of the simulation. Unlike the simulations in
\citet{dev02,dev03,hir04,kro05,haw06,bec08} and \citet{mck04,mck05}
that considered a finite toroidal pool of gas, threaded by poloidal
loops of like orientation, \citet{igu08} considered an endless
supply of gas from the outer boundary. This is a major fundamental
distinction from the poloidal loop simulations. In the poloidal loop
simulations, a finite amount of flux reaches the accretion vortex.
The vortex or funnel fills with the leading edges of the loops of
like orientation. The magnetosphere in the vortex is then only
modestly perturbed by additional vertical flux of random orientation
created by the MRI dynamo within the turbulent accretion flow
\citep{bec08}. By contrast, in the simulation presented in
\citet{igu08}, a continuous supply of vertical flux approaches the
accretion vortex all of the same orientation, even after the vortex
is completely saturated with flux. The question is what does the
numerical system do with the continual pile up of excess flux near
the vortex in a frozen-in accreting plasma? In the 2-D simulations
of \citet{igu08}, a time-depended behavior of the flow with two
different field line topologies resulted near the black hole as
shown in figure 1 (we describe these simulations in more detail in
section 3).
\begin{figure*}
\includegraphics[width=150 mm]{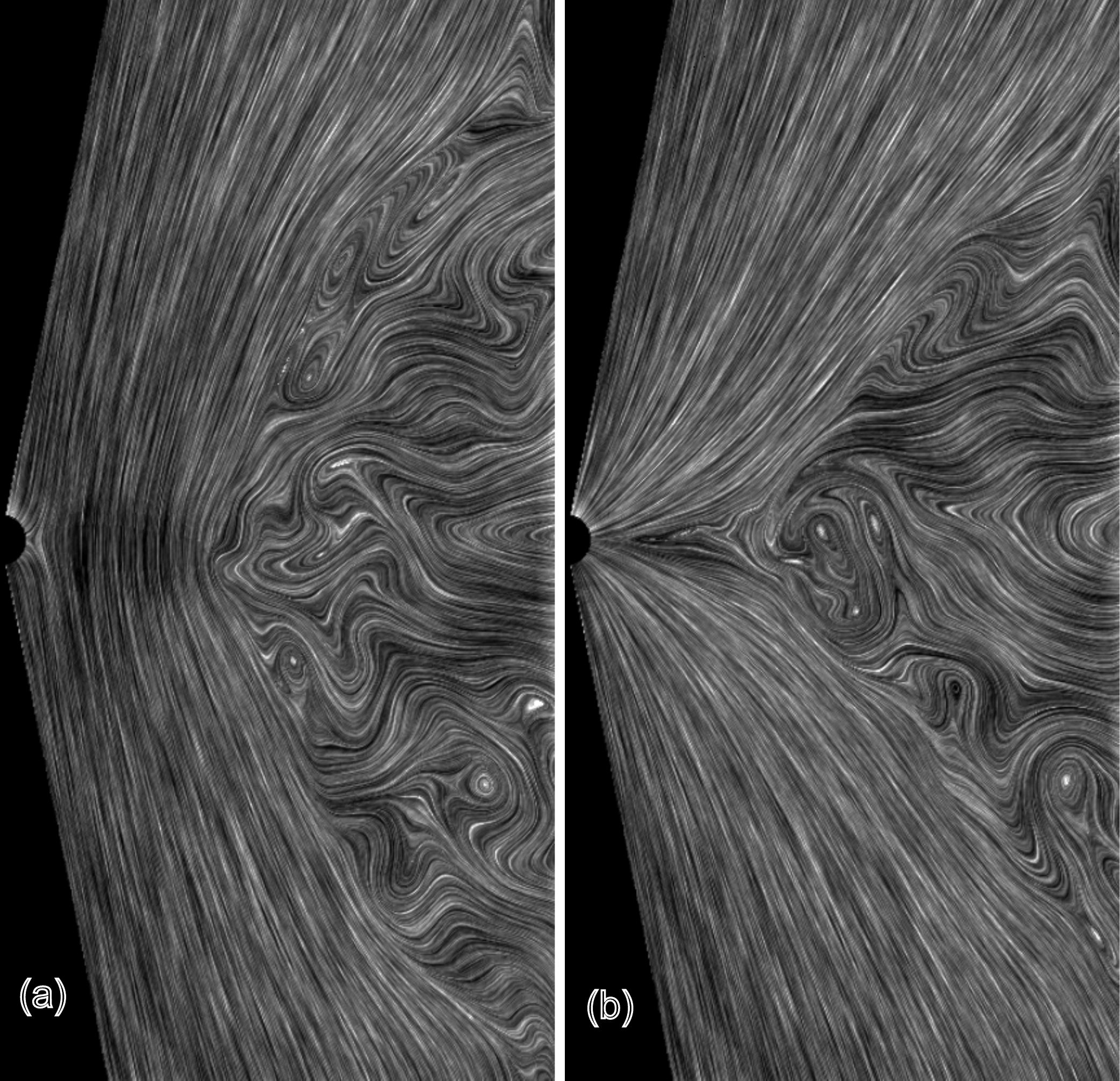}
 \caption{The poloidal magnetic field in the 2-D simulation of \citet{igu08}. The figure is a
 magnification of the inner region of figures 2a and 2b from \citet{igu08}. The topology of the
 right
frame, figure 1b, is the "hourglass" shape that has very little
vertical flux penetrating the equatorial plane near the event
horizon. The gas is free to plunge into the black hole radially. The
left frame, figure 1a, represents the magnetically arrested
accretion state in which the preponderance of flux near the black
hole penetrates the equatorial plane.}
 \end{figure*}
\par Another line of 2-D research was initiated in \citet{kom04}. He
explored the l=1, m=0, vacuum field solution of Maxwell's equations
on a black hole background (known as the Wald solution) after it was
filled with a diffuse plasma. The outer boundary condition was Bondi
accretion. This problem is relevant because the initial state has
vertical flux through the equatorial plane just outside the event
horizon. It was found that initially the accreting low density
plasma accumulated in an arrested state in the equatorial plane.
However, eventually, as the arrested gas continued to accumulate. it
achieved enough inertia to drag the vertical field radially inward
in the equatorial plane into the event horizon making an "hourglass"
topology (similar to figure 1b). There was no vertical flux through
the equatorial plane, near the black hole. This was a similar
topology to what was observed generally in the thick tori accretion
models of \citet{dev02,dev03,hir04,kro05,bec08} and
\citet{mck04,mck05}. This led \citet{kom07} to consider this as a
general topology for black hole magnetospheres - approximately
radial flux at the horizon and none through the equatorial plane
near the event horizon.

\par In many respects, the works of \citet{igu08} and
\citet{kom04,kom07} have crystalized the discussion of vertical flux
near black holes into a simple topological dichotomy: is the
astrophysical state an "hourglass" topology (right frame of figure
1, figure 1b) or a magnetically arrested topology (the left frame of
figure 1, figure 1a)? There is numerical support for both the
hourglass and the vertical topology within the literature. This
article is not one to proclaim which numerical simulation is the
"one" that represents nature. This could not be demonstrated with
anything that has been done numerically or is known theoretically
about the physical state of accreting quasar plasma. All we can say
is that simple approximations to the actual physical system were
evolved with some numerical codes and this is what occurred.

\section{Long Term Flux Accretion in the Pseudo-Newtonian Potential}

In this section, we consider the data from the previously published
simulations in \citet{igu08} to lay the groundwork for demonstrating
the close similarity between the types of accretion flows found in
\citet{igu08} and in fully relativistic simulation KDJ that is
described in section 4. For the purposes of this paper, we reanalyze
the data to explicitly show the vertical magnetic field evolution in
the 3-D simulations. Previously, only plots pertaining to the
density evolution were published. We also explicitly demonstrate the
connection between the evolution of strongly magnetized islands and
density variations within the 3-D flows near the black hole. A
connection was mentioned in \citet{igu08}, but never shown in
detail. These are crucial details that we will need to make a
comparison and contrast to an independent 3-D numerical simulation
that we analyze in section 4. The fact that similar magnetic islands
exist in two different simulations enhances the robustness of our
analysis.

\par The simulations in \citet{igu08}
expanded on the work of \citet{igu03} by studying radiatively
inefficient accretion disks with poloidal magnetic fields in more
detail. These works simulated accretion flows around a
Schwarzschild black hole using the equations of
non-relativistic ideal MHD (\citet{lan87}),
in which the black hole gravity is approximated using a pseudo-Newtonian potential
(\citet{pac80}). A major difference between
the numerical methods of \citet{igu03,igu08} and the simulations
that were presented in \citet{dev02,dev03,hir04,kro05,haw06,bec08}
and \citet{mck04,mck05} is that they allow for the astrophysical
meaningful condition that material is continuously injected into the
system from a distant injection surface, $R_{inj}$, as opposed to
evolving a finite pool of material that was deposited in the initial
state.

\par The simulations in \citet{igu08} are started in 2D, assuming
axial symmetry, with an injection of mass with the Keplerian angular momentum
in a slender torus, which
is located in the equatorial plane at $R_{inj}=210\, R_{g}$, just
inside the outer calculational boundary $R_{out}= 220\, R_{g}$
(where $R_g=2GM/c^2$ is the gravitational radius of the black hole).
Initially there is no
magnetic flux and the injected mass forms a steady (non-accreting)
thick torus. The MHD portion of the simulations are started at $t=0$ from this steady
torus by initiating the injection of a poloidal magnetic field at $R_{inj}$.
The method is the same as
in \citet{igu03}. In each time step, $i$, field is introduced into
the slender torus by adjusting the azimuthal component of the vector
potential (all other components are set to zero) by
$A^{i}_{\phi} = A^{i-1}_{\phi} + \sqrt{8\pi(\Delta\rho)
c_{s}^{2}/\beta_{\rm inj}} \Delta$,
where $(\Delta\rho)$ is the increase in density in the torus during
the time step due to the injection of matter, $c_{s}$ is the sound
speed and $\Delta$ is the grid spacing. The strength of the
injected field can be limited by setting the minimum $\beta_{\rm
inj}$, which is the ratio of the gas pressure to the magnetic
pressure at $R_{\rm inj}$. Large scale magnetic stresses
within the injected material allows for a redistribution of angular momentum. Thus,
some of the injected matter loses angular momentum and accretes
toward the black hole and some matter also flows out of the
computational domain thereby removing excess angular momentum. The
entire volume of the thick torus is filled by the field during about
one orbital period, $t_{\rm orb}$, estimated at $R_{\rm inj}$. At
this moment, $t\simeq t_{orb}$, the formation of accretion flow
begins as a result of redistribution of the angular momentum in the
torus due to Maxwell stresses.
\par Various simulations of accreting vertical flux were run in
\citet{igu08} in which the only variable was $\beta_{inj}$, the
relative strength of vertical magnetic field that was injected.
It should be noted that in the absence of radiative losses the
accretion flow structure is independent of the assumed
mass accretion rate and the dependence on the black hole mass $M$
come only through the spatial scales $R \propto R_{\rm g}$.
The basic trend was as follows, the smaller $\beta_{inj}$ (the stronger the
injected field) the weaker the MRI (turbulence) and the larger the
magnetic stresses (i.e, the faster angular momentum is removed
outward from the accretion flow and the more rapidly the plasma
accretes toward the black hole). The simulation known as model B was
investigated in the most detail in \citet{igu08} and is defined by
$\beta_{inj} = 100$. We will expand on the discussion of
\citet{igu08} in the context of the present topic in this section.

\subsection{The 2-D Results}

The initial evolution in the models in \citet{igu08} was simulated
in 2D. This allows one to consider longer evolution times in
comparison to those that can be obtained in 3D simulations, because
of the larger requirements for computational resources in the latter
case. The 3D simulations are initiated from developed axisymmetric
models. An extended duration 2-D run of model B was performed in
parallel with the 3-D run. The relevant details are captured in
figures 1a and 1b. The figure samples two data dumps, at $t=5.1153$
and $t=5.1458$ (left and right frames), given in units of the
Keplerian orbital period at $R_{inj}= 210 R_{g}$. The magnetic field
lines are produced in figure 1, by the line integral convolution
method \citep{cab93}. The frame on the left (figure 1a) has a large
patch of vertical flux through the equatorial plane that extends
almost all the way to the inner circular calculational boundary,
$R_{\rm in}=2R_g$. The field deviates from purely vertical near the
equatorial plane extremely close to the boundary in order to meet
the imposed boundary condition of no transverse magnetic field at
the inner calculational boundary, $R_{in}$.  Even with this boundary
condition, the field is vertical at the equator for $R-R_{in} > 0.75
R_{g}$. In this phase, the accretion flow is halted by the patch of
strong vertical flux. This is the 2-D version of a magnetically
arrested disk \citet{igu03,igu08}.
\begin{figure}
\epsscale{0.93}\plotone{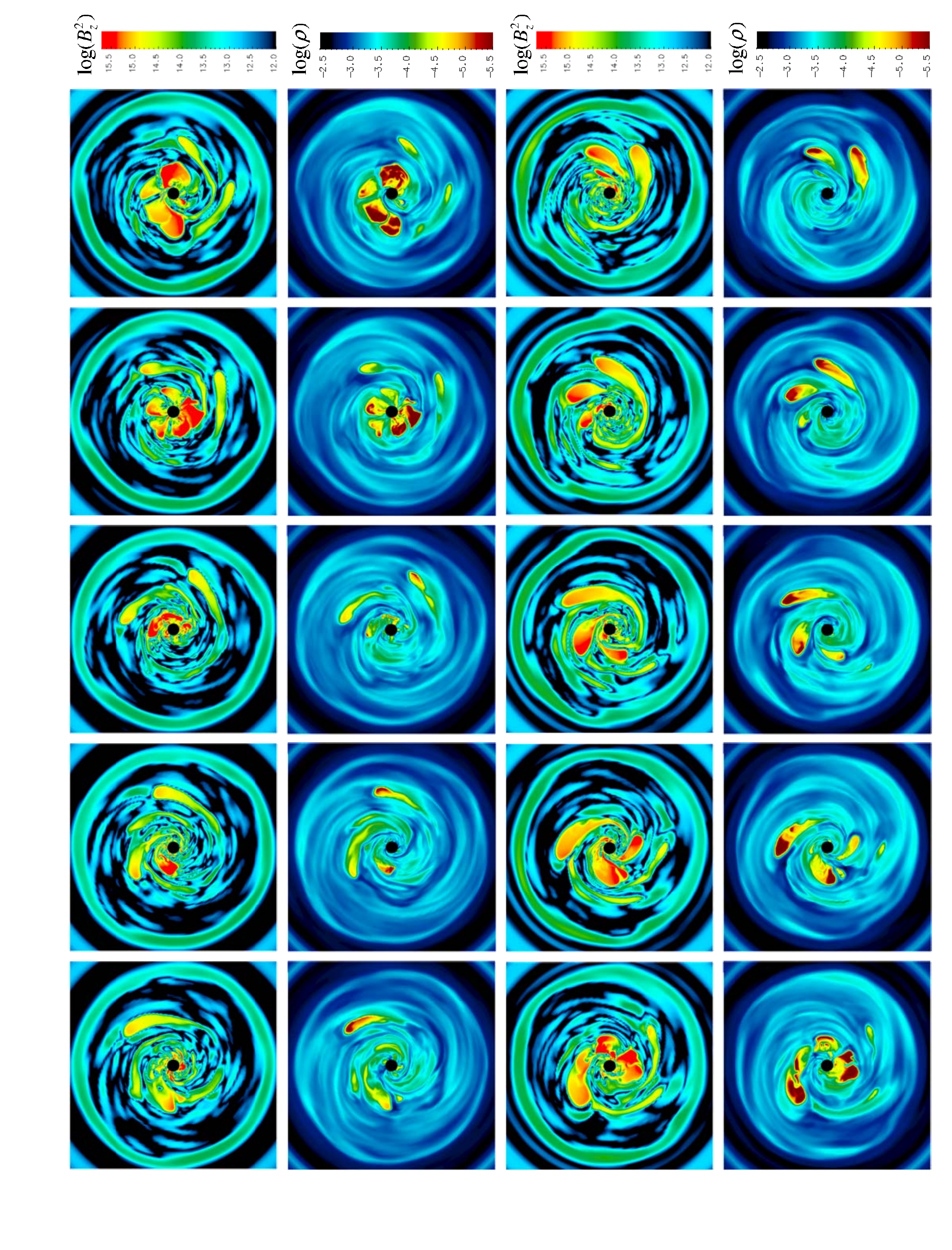}

 \caption{10 consecutive time snapshots of the poloidal magnetic field and density in model B of \citet{igu08}. The left column and the
 second column are simultaneous consecutive time snapshots in the equatorial plane of  $B_{z}^{2}$  and density, respectively.
 Similarly for columns 3 and 4. The time increases from bottom to top and column 1 precedes column 3 (see text for details). The second and fourth columns are
 extracted from movie 1 that is on-line.}
 \end{figure}

\par The magnetically arrested portion displayed in figure 1a keeps
accumulating mass and field as the accretion from large distances
proceeds. However, the radial force balance between the trapped
magnetic flux and the incoming gas is Rayleigh-Taylor unstable, but
in 2-D there is the assumption of axisymmetry, so ingoing flutes of
plasma can not slide by on the side (the ``side" is the the $\phi$
direction) of islands of buoyant magnetic flux as it would in a 3-D
Rayleigh Taylor interface. In 2-D, the mass density builds up until
the increased attraction towards the gravitational center drags the
field and "stacked up" frozen-in plasma inward radially as in figure
1b. When the field attains this "hour-glass" topology, the plasma is
allowed to accrete freely into the black hole in the radial
direction by just sliding along the field lines. Recall that the
same situation occurred in the 2-D simulation of \citet{kom04} that
was discussed in the previous section.
\par The plasma drains down the "hour-glass" rapidly. Once the
density is sufficiently low, the field topology actually reverts
back to the magnetically arrested topology of vertical flux near the
inner boundary. A new magnetically arrested epoch is begun.
Actually, the system spends almost all of its time in the
magnetically arrested state as depicted in figure 1a and only goes
into the the "hour-glass" for brief instances separated by large
time intervals. Compare this to the 2-D simulations of
\citet{kom07}, in which the field is in a pure hour-glass topology
all of the time. In particular, figure 3 of \citet{kom07} shows an
accretion disk flowing through the "cusp" of the "hour-glass" near
the equatorial plane. By contrast, model-B of \citet{igu08} is only
rarely in this "hour-glass" field topology and is in the
magnetically arrested state the vast majority of the time. Recall,
the fundamentally different assumptions made in these two articles.
The \citet{kom07} accretion flow contains a finite amount of
poloidal flux and there is no large reservoir of flux that accretes
from large distances. By contrast, \cite{igu08} assumes an eternal
supply of gas and vertical field that can accrete from large
distances.
\subsection {The 3-D Results}
 Next we
explore the more physically relevant 3-D portion of the simulations.
Neither state in figure 1a nor figure 1b is achieved. In 2-D the two
states in figure 1 are mutually exclusive in a temporal sense. In
3-D, the two states present in the 2-D flow seem to coexist
simultaneously within the same flow. Not surprisingly, this leads to
some nontrivial dynamics that we study by means of the series of
snapshots presented in figure 2. Figure 2 represents 10 consecutive
data dumps separated by $\Delta t = 17.8 M$, where $t$ is in
geometrized units, and plots two quantities at each corresponding
time, the logarithm of the density, $\rho$ and the logarithm of the
square of the vertical field, $B_{z}^{2}$, in the equatorial plane.
Figure 2 contains a wealth of information that is far too detailed
for a figure caption. From left to right, columns 1 and 3 are false
color plots of $\log (B_{z}^{2})$.  Time progresses upward in each
column and the column 1 precedes column 3. Columns 2 and 4 are false
color contour plots of $\log (\rho)$. These two columns are
snapshots of the animation, movie 1, that is available in the
on-line material. Note that the plasma rotates in the clockwise
direction in figure 2.

\par The snapshots in figure 2 were chosen to elucidate the
evolution of magnetic flux build-up in the inner regions of the
accretion flow. Consider frame 1 in the first column. Blue
represents a small vertical magnetic field component and red
indicates a large vertical field component at the equator. There are
actually small regions of strong vertical field just outside the
inner calculational boundary. However, most of the vertical field is
about an order of magnitude weaker in the remainder of the inner
region. In frames 2 and 3 (from the bottom) in column 1, there is a
definite accumulation of vertical flux near the inner boundary. The
peak vertical flux close to the inner boundary is in frame 4. After
this, the large patch of strong vertical field starts migrating away
from the inner calculational boundary in frame 5.

\par Now consider the the second column. In this plot, the color
scheme is reversed, blue is high density and red is low density. In
the first three frames of column 2 (going from bottom to top) the
density is high, in code units, $\rho \sim 10^{-3}\, -\, 10^{-4}$
within the evolving magnetic islands. After the peak flux is
accumulated near the inner calculational boundary in frame 4, the
density drops precipitously, $\rho \sim 10^{-5}\, -\, 10^{-5.5}$
within the magnetic islands. This is the equivalent to a local
region of a magnetically arrested accretion flow. Very little mass
penetrates the strong magnetic pressure barrier. In 2-D, this
phenomena is global in figure 1a, but in 3-D it exists in localized
regions with most of the plasma accreting along pathways the flow
around the edges of these magnetic islands. In this context, we note
that throughout the simulation, the vertical flux and density
variations are not axisymmetric, but dominated by spiral modes with
azimuthal wavenumbers, $1\le m\la 5$ \citep{igu08}.

\par Regions of strong vertical flux ($\beta \sim 0.01$) exist extremely close to the inner
calculational boundary, much closer than in the 2-D case presented
in figure 1a. The local magnetically arrested areas associated with
the enhanced vertical field strength are inherently Rayleigh-Taylor
unstable. The strong magnetic islands in frame 4 of column 1 begin
to migrate slowly outward in frames 5 and 6 (the bottom frame of
column 3). Furthermore, the large magnetic pressure in the magnetic
islands makes them "over-pressurized" relative to regions outside the islands.
The pressure begins to spread the islands apart
in frame 5. Since the magnetic islands are low density they are
weakly affected by gravity. By contrast, the dense accretion flow is
strongly attracted to the central black hole. The magnetic islands
therefore become buoyant within the dense accretion flow under the
influence of the gravitational field. This makes them spiral outward
in frames 6 - 9, creating a strong shear relative to the enveloping
accretion flow. As the buoyant flux tubes spiral outward, magnetic
flux is constantly diffusing (by numerical diffusion working in
consort with the shear forces) into the surrounding accretion flow.
Consequently, as the buoyant flux tubes spiral outward the field
strength weakens considerably as the flux mixes back into the
accretion flow in frames 6-10. This temporal evolution is repeated quasi-periodically. Similar
cycles of vertical flux accretion/expulsion near black holes was
anticipated in \citet{pun90} and is discussed in Chapter 8 of
\citet{pun01}.

\subsection{Summary of Relevant Details}
We summarize the key results from model-B of \citet{igu08}.
\begin{enumerate}
\item The time evolution of the vertical magnetic flux in 3-D is
qualitatively different from that in 2-D .
\item In 3-D, strong patches of vertical magnetic flux
form episodically and are located in the innermost regions of the accretion flow.
\item Plasma behaves as a two component system. There are the magnetic islands
(strong magnetic field, low density regions) that tend to exclude
the accretion flow (the 3-D version of a magnetically arrested
accretion flow exists only locally). The second component is the
bulk accretion flow in the form of spiral streams that "swims around" the magnetic islands by
interchange instability.
\item The magnetic islands are not time stationary. They become
buoyant and the flux is redeposited back into the outer accretion
flow.
\item The spiral flow twists the vertical flux
forming collimated bipolar Poynting outflows, or jets, that
extract $\sim 1$\% of the binding energy of the accretion
flow.
\end {enumerate}

\section{Vertical Magnetic Flux Near Rapidly Rotating Black Holes} In this section, we turn our
attention to vertical magnetic flux evolution near a rapidly
rotating black hole in the Kerr spacetime. The metric of the Kerr
spacetime (that of a rotating uncharged black hole) in
    Boyer-Lindquist coordinates, $g_{\mu\nu}$, is given by
    the line element
    \begin{eqnarray}
        && d s^{2} \equiv
        g_{\mu\nu}\, dx^{\mu} dx^{\nu}=-\left
(1-\frac{2Mr}{\rho^{2}}\right)
        d t^{2}
        +\rho^{2} d\theta^{2}\nonumber\\
&& +\left (\frac{\rho^{2}}{\Delta}\right) dr^{2}
     -\frac{4Mra}{\rho^{2}}\sin^{2}\theta\nonumber\\
&&  d\phi \,
        dt+\left [(r^{2}+a^{2})+\frac{2Mra^{2}}{\rho^{2}}\sin^{2}
\theta
        \right ] \sin^{2} \theta \, d\phi^{2} \; ,\\
        &&\rho^{2}=r^{2}+a^{2}\cos^{2}\theta ,\\
        &&\Delta = r^{2}-2Mr+a^{2} \equiv \left
        (r-r_{_{+}})(r-r_{_{-}} \right )\;.
        \end{eqnarray}
There are two event horizons given by the roots of the equation
        $\Delta=0$. The outer horizon at $r_{_{+}}$ is of physical
interest
        \begin{eqnarray}
&& r_{_{+}}=M+\sqrt{M^{2}-a^{2}} \; .
            \end{eqnarray}
The stationary limit surface, at $r=r_s$ occurs when $\partial /
\partial t$ changes from a timelike vector field to a spacelike
vector field.
\begin{equation}
r_s = M + \sqrt{ M^2 - a^2 \cos^2 \theta} \; .
\end{equation}
This change happens when $g_{tt}$ switches sign. The region in which
$g_{tt}>0$ is the ergosphere. The outer boundary of the ergosphere
is the stationary limit surface.
\par The lessons of the last
section tell us that the time evolution of vertical magnetic flux
near a black hole in 2-D and 3-D simulations can differ at the most
fundamental levels. Thus, if we want to truly understand the
properties of vertical flux distributions near black holes it is
wise to work in 3-D. The only long term, high resolution, 3-D MHD
numerical simulations in the Kerr spacetime are those developed in
\citet{dev02,dev03,hir04,kro05,haw06,bec08}. There is no radiation term in the time evolution
equations. Thus, as heat is generated, it can flow and be converted
to gas pressure, but not to a flux of radiation. Thus, this is very
different than a quasar.
\par The reader should
refer to \citet{dev02,dev03,hir04,kro05,haw06,bec08} to get more
details on the many simulations that have been performed with this
code over the years. This discussion is no more than a brief review.
The initial state is a thick torus of gas in equilibrium that is
threaded by concentric loops of weak magnetic flux that foliate the
surfaces of constant pressure. The magnetic loops are twisted
azimuthally by the differentially rotating gas. This creates
significant magnetic stress that transfers angular momentum outward
in the gas, initiating a strong inflow that is permeated with MRI driven turbulence. The
end result is that after t = a few hundred M (in geometrized units),
accreted poloidal magnetic flux gets trapped in the accretion vortex
or funnel (with an opening angle of $\sim 60^{\circ}$ at the horizon
tapering to $ \sim 30^{\circ} - 35^{\circ}$ at $r > 20 M $). The
orientation of the field in the funnel is the same as the leading
edge of the poloidal loops in the initial state. The outer edge of
the poloidal loops are initially driven outwards and never accrete
within the lifetime of the simulation. The strong transients die off
by t = 2000 M, so the late time data dumps are the most physically
relevant \cite{haw06}.

\par Numerically, the problem is formulated on a grid that is 192 x
192 x 64, spanning $ r_{in} < r < 120M $, $8.1^{\circ}< \theta <
171.9^{\circ}$ and $0 < \phi < 90^{\circ}$. Excision of the thin
polar cone restricts the time steps to be not too small in the used
spherical coordinates (the Courant condition) and greatly reduces
the amount of computational resources required. The inner
calculational boundary, $r_{in}$, is located close to, but just
outside of the event horizon, $r_{+}$, where the Boyer-Lindquist
coordinates are singular. The $\phi$ boundary condition is periodic
and the $\theta$ boundary conditions are reflective. Zero-gradient
boundary conditions are employed on the radial boundaries, where the
contents of the active zones are copied into the neighboring ghost
zones. MHD waves propagate slower than the speed of light, therefore
the gravitational redshift creates a magneto-sonic critical surface
outside of $r_{+}$ from which no MHD wave can traverse in the
outward direction \cite{pun01}. The philosophy was to choose
$r_{in}$ to lie inside the magneto-sonic critical surface, thereby
isolating it from the calculational grid. There are also steep
gradients in the metric derived quantities as $r_{+}$ is approached.
This is handled by increasing the resolution of the grid near
$r_{in}$ with a "cosh" distribution of radial nodes.
\par We caution the reader that
the results of these types of simulations are highly sensitive to
the initial conditions that are imposed. Of all the loop
configurations that are tried, only poloidal loops of the same
orientation in a dipolar configuration produce strong Poynting jets.
In \citet{bec08}, it was shown that if the loops are in a
quadrupolar orientation in the initial state, the Poynting jet power
coming from near the black hole is reduced to about 1\% of the value
obtained from the common orientation dipolar loops. If the loops are
toroidal in the initial state, the Poynting jet power coming from
near the black hole is reduced to about 0.1\% of the value obtained
from the common orientation dipolar loops. Finally, \citet{bec08},
showed that if the initial state of the torus is threaded by dipolar
loops with more than one orientation, a strong Poynting jet only
forms if a coherent packet of loops of like orientation accretes.
\par One of the highest spin simulations in this family is from
\citet{haw06} and is known as KDJ, $a/M=0.99$.  The last three time
steps of KDJ with data collection were generously shared by J.
Krolik and J. Hawley. Expressing the mass in geometrized units, the
data dumps were performed at t = 9840 M, t = 9920 M and t = 10,000
M. This complex simulation was crudely studied in
\citet{pun07,pun08}. using 2-D visualization techniques. Due to the
complex inhomogeneous nature the averages over azimuth that were
employed did not yield a clear picture of the field configuration
near the black hole. The purpose of this section is to use new 3-D
visualizations techniques that allow us to clearly understand the
varied vertical flux structures that form near the rapidly rotating
black hole in KDJ. It was pointed out in \cite{pun01}, the
unfortunate circumstance that all the plots from KDJ in
\citet{pun07,pun08} were displayed upside down. The north and south
poles were reversed in all those plots, this is remedied here. The
main emphasis in this study is to associate certain vertical field
line topologies with the strong Poynting flux flares that emerge
from the ergosphere in KDJ.
\begin{figure*}

\includegraphics[width=110 mm]{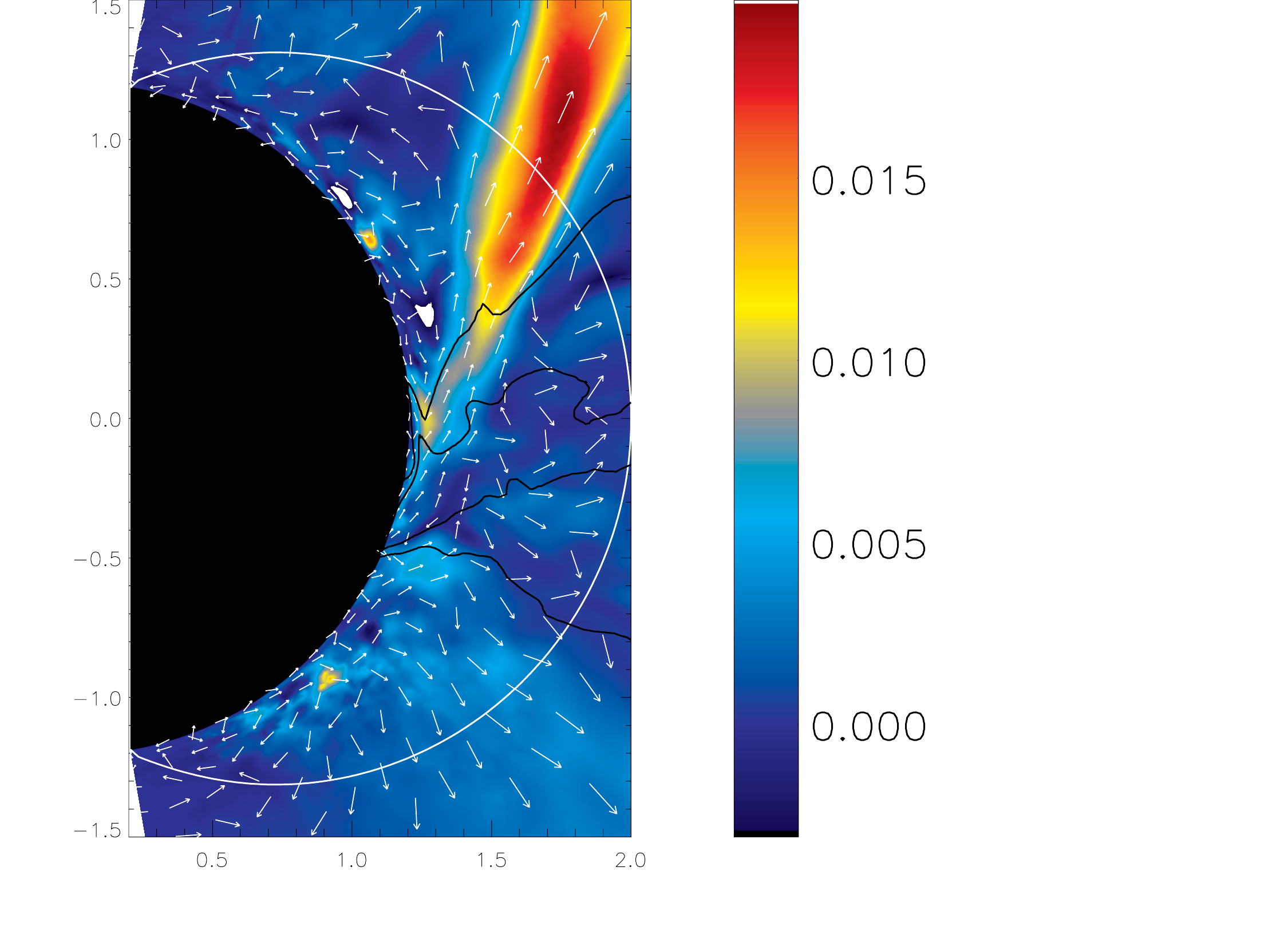}
\hspace{-2.0cm}
\includegraphics[width=110 mm]{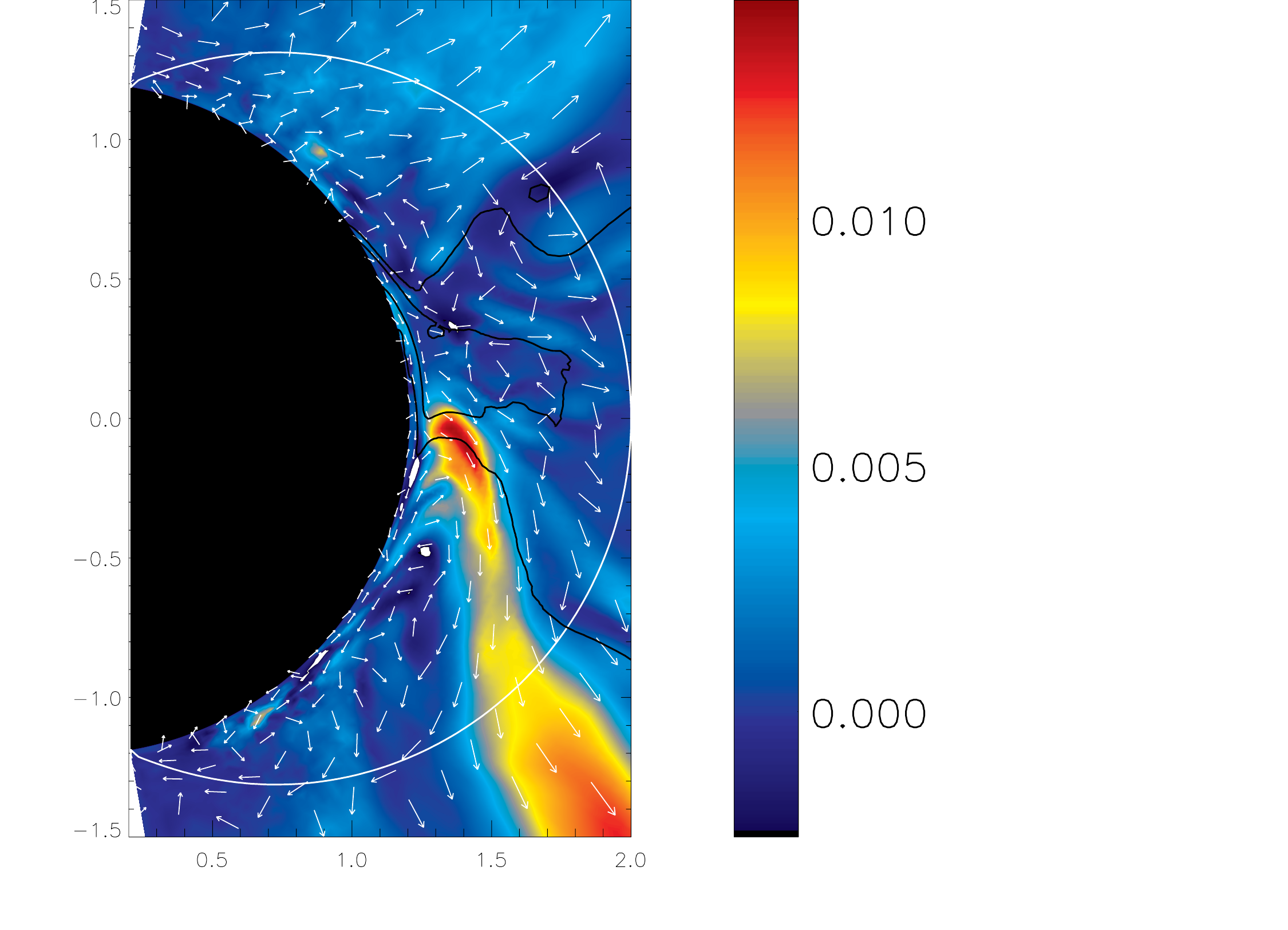}

\includegraphics[width=110 mm]{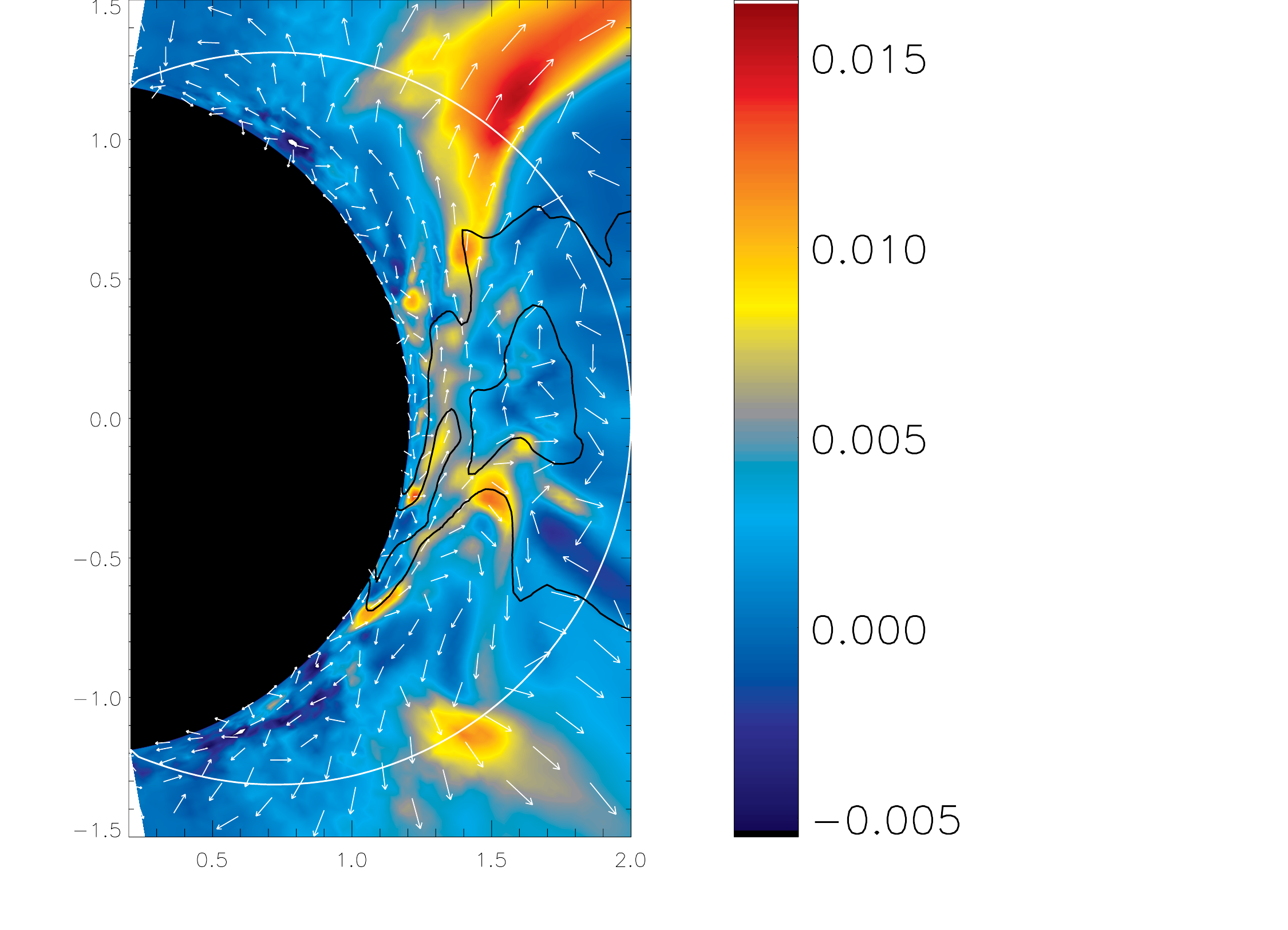}
 \caption{A false color plot of $S^{r}$ averaged over azimuth at three time steps of KDJ, t = 9840 M (top left),
 t = 9920 M (top right) and t = 10000 M (bottom row). The relative units
(based on code variables) are in a color bar to right of each plot
for comparison of magnitudes between the plots. Superimposed are
white arrows that represent the direction of the
 azimuthally averaged poloidal component of the Poynting flux, $\mathbf{S}^{P}$. The length does not represent the strength of the
 poloidal Poynting flux, but is proportional to the grid spacing. The white contour is the stationary limit surface. The black contours on the
plots are the Boyer-Lindquist evaluated density, scaled from the
peak value within the frame at relative levels 0.5 and 0.1. The
inside of the inner calculational boundary (r=1.203 M) is black. The
calculational boundary near the poles is at $8.1^{\circ}$ and
$171.9^{\circ}$. There is no data clipping, so saturated regions
appear white.}
\end{figure*}
\subsection{The Poynting Jet in KDJ}
The conservation of global, redshifted, or equivalently the
Boyer-lindquist coordinate evaluated energy flux, defined in terms
of the stress-energy tensor, is shown in \citet{pun07} to simply
reduce to
\begin{eqnarray}
&& \partial(\sqrt{-g}\, T_{t}^{\, \nu})/\partial(x^{\nu})=0\; .
\end{eqnarray}
The four-momentum $-T_{t}^{\, \nu}$ has two components: one from the
fluid, $-(T_{t}^{\, \nu})_{\mathrm{fluid}}$, and one from the
electromagnetic field, $-(T_{t}^{\, \nu})_{\mathrm{EM}}$. The
quantity $g = -(r^{2} + a^{2} \cos^{2}{\theta})^{2}\sin^{2}{\theta}$
is the determinant of the metric. The integral form of the
conservation law arises from the trivial integration over Boyer -
Lindquist coordinates of the partial differential expression
\citep{thp86}. It follows that the the redshifted Poynting flux
vector can be defined in Boyer-Lindquist coordinates as
\begin{eqnarray}
\mathbf{S}= (S^{r},\; S^{\theta},\; S^{\phi})=(-\sqrt{-g}\,
(T_{t}^{\, r})_{\mathrm{EM}},\; -\sqrt{-g}\, (T_{t}^{\,
\theta})_{\mathrm{EM}},\; -\sqrt{-g}\, (T_{t}^{\,
\phi})_{\mathrm{EM}} )\; .
\end{eqnarray}
\par After averaging over azimuth, we use the poloidal projection of
the Poynting vector, $\mathbf{S}^{P}$, to display the propagation of
electromagnetic energy flux out of the ergosphere in KDJ. Consider
figure 3, which is a false color plot of the azimuthal average of
$S^{r}$ at the three time steps. Notice that in each time frame, the
flow of $S^{r}$ across the white boundary (the stationary limit) is
dominated by narrow channels that seem to initiate suspended outside
the event horizon ($r_{+} = 1.175$ M) and the inner calculation
boundary (shaded black) at r= 1.203 M. In order to explain the
origin of these energy beams, we overlay the direction of the
poloidal Poynting vector in the each of the frames in figure 3. The
white arrows represent the poloidal direction of the azimuthally
averaged poloidal component of the Poynting flux, $\mathbf{S}^{P}$.
The length of the arrows is proportional to the grid spacing and is
not related to the magnitude of $\mathbf{S}^{P}$. The black contours
on the plots are of the Boyer-Lindquist evaluated density, scaled
from the peak value within the frame at relative levels of 0.5 and
0.1. The integral curves that are approximated by the white arrows
show that the primary source of $S^{r}$ is Poynting flux emerging
from the accretion flow near the equatorial plane of the ergosphere
in all three frames. The direction of the white arrows indicate that
$\mathbf{S}$ has a large $S^{\theta}$ component at its point of
origin near the equatorial plane. Note that the energy conservation
law in equation (4.6) is dominated by the electromagnetic terms for
the energy flux emitted from the ergosphere (the mechanical energy
flux is $\sim$ 1\% - 5\% of the electromagnetic energy flux emerging
from the ergosphere). In \citet{pun07}, it was shown that there is a
strong vertical magnetic field component at these locations as well.
In this article, we actually create 3-D plots of the field lines to
show the connection between field line topology and the creation of
ergospheric Poynting flux in KDJ. We explore the possible vertical
field configurations by studying each time slice individually.
\par The data presented in figure 3 is used to determine the relevant field
lines to plot in 3-D that are required to elucidate the origin of
the dominant component of the Poynting flux emerging from the
ergosphere in KDJ, namely from the accretion flow near the
equatorial plane. If we naively integrate field lines from random
points, one gets a clutter of field lines that wrap around the black
hole many times. It is impossible to extract the source of the
strong Poynting jet from inner edge of the ergospheric accretion
flow in the midst of this clutter. Our desire is to find the field
lines that support this Poynting jet. We are not interested in
plotting the field lines that seem to spiral endlessly in the
equatorial plane and never connect to a jet. This will just mask the
base of the ergospheric Poynting jet. Furthermore, the field lines
that support the weaker Poynting flux contribution associated with
the horizon magnetosphere (the Blandford-Znajek) type field lines
are also not the focus of this study. For the interested reader, we
note that these field lines have already been plotted for similar
simulations in \citet{hir04}. The exposition of these field lines
would also clutter our view of the source of the dominant component
of the ergospheric Poynting jet. Our final plots will have a wealth
of detail as is. Thusly motivated, based on figure 3, we restricted
the latitude of the starting points of field line integration to
$\pm 4^{\circ}$ of the equatorial plane and the interval, $1.3M < r
< 1.6 M$. This choice of integration starting points captures the
poloidal magnetic flux that is contained within the dominant
component of the ergospheric Poynting jet.

\subsection{The time-slice t =9840 M}
This section begins our detailed study of the vertical flux in KDJ
using 3-D visualization. The data from each time slice is presented
in chronological order. The first time slice to be presented is at t
= 9840 M. This time slice shows most of the features that were
observed in the cumulative data from KDJ. It is perhaps the richest
in terms of the variety of magnetic features and it is fortuitous
starting point for our work.
\begin{figure}

\includegraphics[width=160 mm]{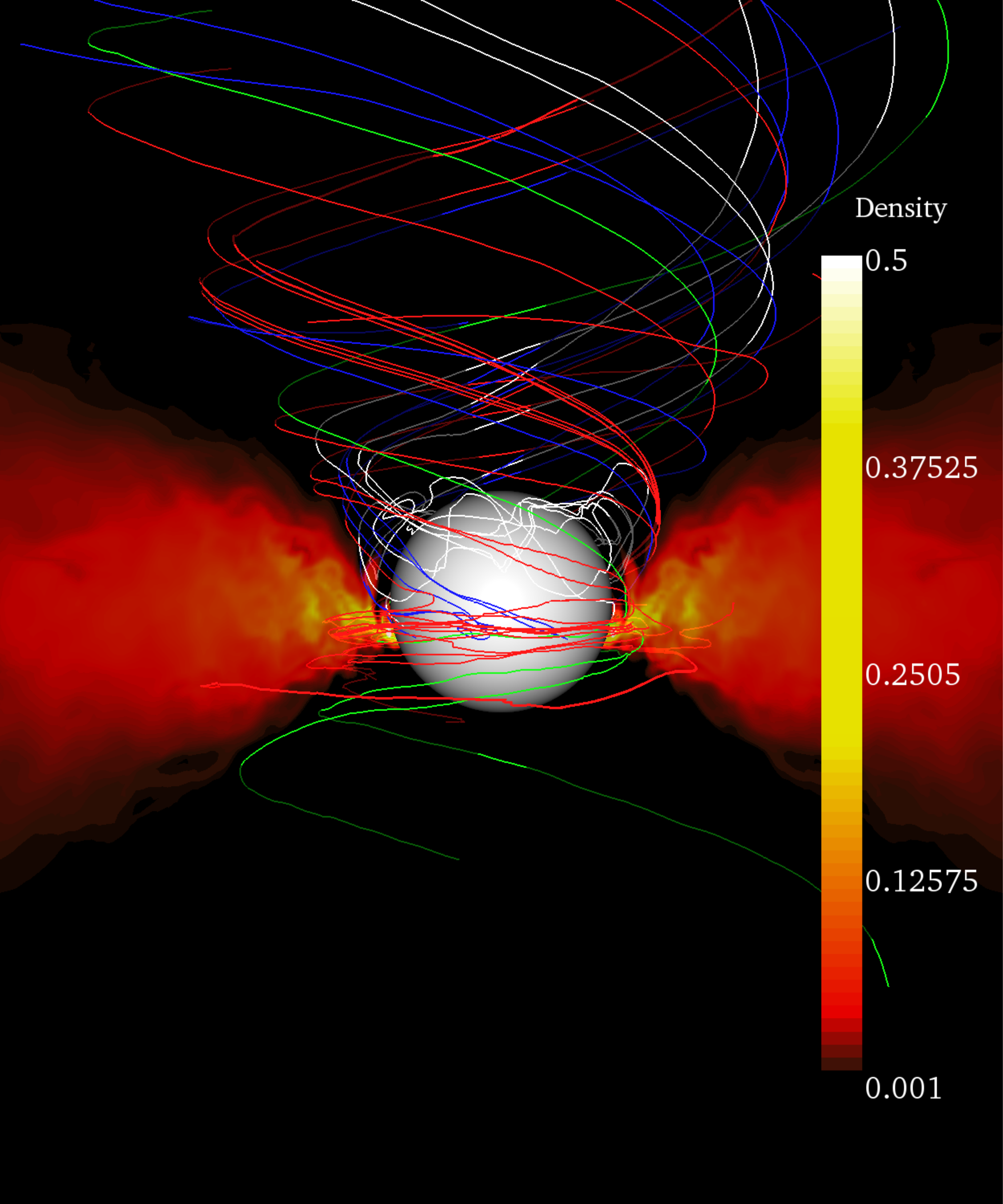}

 \caption{Magnetic field lines superimposed on the false color plot of Boyer-Lindquist
evaluated density in a cross-section within the plane that is
perpendicular to the line of sight. The values of the density are
designated by the color bar in code units on the right hand side of
the figure. The inner calculational boundary is shaded grayish-white
for maximum contrast. The jet emerges from inner edge of the
equatorial accretion flow.}
\end{figure}

\begin{figure}

\includegraphics[width=150 mm]{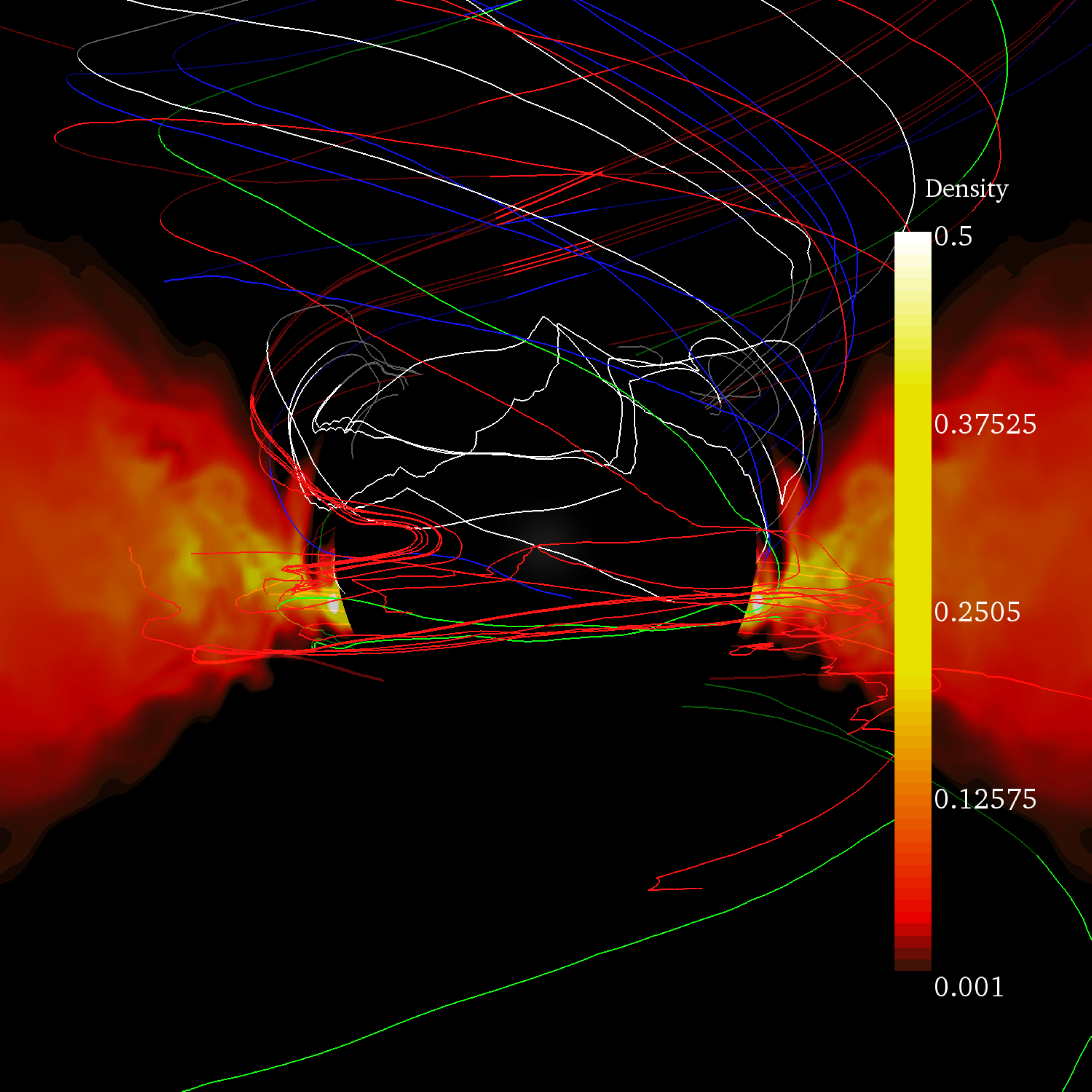}

 \caption{Magnetic field lines superimposed on the false color plot of Boyer-Lindquist
evaluated density in a cross-section defined by the plane that is
perpendicular to the line of sight. The values of the density are
designated by the color bar in code units on the right hand side of
the figure. The black hole is black. The jet emerges from inner edge
of the equatorial accretion flow. In this figure, the camera in
Paraview is rotated $180^{\circ}$ in azimuth relative to figure 4.
It is a closeup that highlights the interaction of the field lines
and the accreting gas seen in cross-section. The inner calculational
boundary is shaded black.}
\end{figure}
\subsubsection{The Poynting
Jet at t =9840 M} Figure 4 is a wide angle view of field lines that
were integrated with Paraview 3.3.0 per the methods in the Appendix
at the data dump corresponding to the Boyer-Lindquist time, t = 9840
M. The Paraview camera in figure 4 is positioned in the equatorial
plane. Figure 5 is a closeup of the same 3-D geometry that exists at
t = 9840 M, viewed along the equatorial plane. The camera in
Paraview is rotated $180^{\circ}$ in azimuth in figure 5 relative to
figure 4. The advantage of the closeup is to highlight the
interaction of the field lines and the accreting gas seen in
cross-section. The advantage of the wide angle view is to highlight
the large scale jet. The backgrounds of figures 4 and 5 are false
color contour maps of a 2-D cross-section of the Boyer-Lindquist
evaluated density in the plane that is perpendicular to the line of
sight. The values of the density are designated by the color bar in
code units on the right hand side of the figure. The interior of the
inner calculational boundary is shaded grayish-white in figure 4 for
maximum contrast. The interior of the inner calculational boundary
is left black in the closeup view presented in figure 5. The density
is chosen with an opacity of $80\%$ so that field lines can be seen
penetrating the accreting gas from the back side to the front.
Consequently, when the field lines are on the back side of the plane
of the figure they appear darker, this makes it easier to visually
trace out the spirals of the magnetic field in the jet. If the
opacity of the density cross-section were chosen to be 100\% then
the field lines would be completely obscured on the back side.
Recall that the polar regions within $8.1^{\circ}$ are excised for
computational efficiency. Thus there is no data (including gas
density) in this region. Consequently, every time a field line on
the back side crosses these polar regions it suddenly becomes as
vivid as the near side part of the field line (there is nothing that
is opaque that blocks our view).
\par The origin of the preponderance of the Poynting flux
at t = 9840 M (the narrow red channel in the top left frame of
figure 3) is almost completely obscured by a tangle of field lines,
many that wrap around endlessly in the equatorial accretion flow.
From the top left frame of figure 3, it is clear that the strong
narrow channel that dominates the northern Poynting jet originates
from near the equatorial plane at $r < 1.55 M$. This is the region
that is relevant to our study and the field lines in figures 4 and 5
were chosen accordingly. In order to produce figures 4 and 5, we
selected 60 starting points for the field line integrations as
follows. First of all, we restricted the latitude of the starting
points of integration to $\pm 4^{\circ}$ of the equatorial plane.
With this restriction on latitude, we chose 40 points at random from
the interval, $1.35M < r < 1.55 M$, and an additional 20 starting
points for integration were chosen randomly from the interval, $1.30
M < r < 1.60 M $. The field lines were integrated in both
directions. The integrations were terminated if the field line
penetrated the computational boundaries. We used 1000L steps per
integration of field lines of length L in units of the geometrized
mass. Increasing this to 10000L steps in the Runge-Kutta integration
method showed only minor differences. Thus, we concluded 1000L steps
were sufficient in the integration routine. Each field line was
integrated 3 times, with lengths, L, of 10M, 20M and 100M. Varying L
on a case by case basis was advantageous for increasing the clarity
of display in the figures. Ideally, having all the lines at 100 M is
best for seeing the jet. But many of the field lines in the jet coil
endlessly at their base within the accretion flow (in particular the
type I lines that are discussed in detail below). The coiling in the
accretion flow blocks off and confuses any details of the other
interactions of the field lines with the accreting gas. Ultimately,
it is just an incomprehensible mess of toroidal coils. For these
types of field lines, we typically chose a length of 20 M for
display purposes. This ended up creating a little more than one
toroidal circuit in the accretion flow (this is why the type I field
lines in some of the wide angle views don't extend as far into the
jet as the other field lines that are integrated to a length of 100
M).
\par Out of the 60 starting points, 22 of the field lines that
were generated connected to the bipolar Poynting jets. Most of the
others were predominantly toroidal in nature and did not connect to
the Poynting jet. Some of these toroidal loops seemed to spiral
endlessly near the black hole. Others passed through the equator and
spiraled outward in the accretion flow into the opposite hemisphere.
Others crossed the equatorial plane and spiraled out into the corona
above the disk into the opposite hemisphere, propagating to large
distances, but never actually merging with the Poynting jet. These
predominantly toroidal field lines obscured the base of the vertical
field lines that did join the Poynting jet, so they were omitted
from figures 4 and 5 since they do not shed light on the topic at
hand, but actually hide interesting details from view. There were
also a few loops that came out of the inner calculational boundary
then intersected the accretion flow and finally closed on the inner
calculation boundary at high latitude. This was only time slice in
which such loops were prevalent.
\par The field lines were color coded in figures 4 and 5 for the discussion of the taxonomy of vertical field lines
in the next subsection. To our surprise, the vertical flux through
the equator is not a single type of field line, but we have
discovered 4 different types of vertical field lines near the
equatorial plane that comprise the large scale jet. The coiling in
the field lines indicates that there is a Boyer-Lindquist evaluated
magnetic field component, $B^{\phi}$, in the compact notation of
equation (A.7). Also, since there is a non-zero pitch angle to the
large scale field, there is a poloidal magnetic field component in
the low density polar region of figure 4, $\mathbf{B^{P}}\equiv
(B^{r},\; B^{\theta})$ in the notation of equation (A.7). Since it
was shown in \citet{pun07} that the field lines at the base of the
jet at t = 9840 M, rotate with an angular velocity that is
approximately that of the event horizon, $\Omega_{F}\approx
\Omega_{H}$, in the Boyer-Lindquist coordinates, there is an
associated motional, poloidal (cross-field) EMF, $\mathbf{E_{P}}\sim
\Omega_{F}(B^{\theta},\; -B^{r})$. Consequently, there is a poloidal
Poynting vector $\mathbf{S^{P}} \sim \mathbf{E_{P}}\times
B^{\phi}\mathbf{e_{\phi}} \sim (\Omega_{F}F_{\phi}^{\;\theta}
F_{\theta r},\; \Omega_{F}F_{\phi}^{\; r} F_{\theta r})$, where the
covariant Boyer-Lindquist basis vector $\mathbf{e_{\phi}}$ is the
coordinate vector field $\partial/\partial\,\phi$ \citep{pun01}.
Thus, the existence of rotating, coiled field lines in the accretion
vortex is equivalent to a Poynting flux jet. Notice that the twisted
field lines in figures 4 and 5 exit the accreting gas at roughly the
same place that the narrow, strong channel of poloidal Poynting flux
emerges from the azimuthally averaged image in the top left frame of
figure 3. These 3-D coiled field lines explain why the white arrows
(Poynting vector) in the top left frame of figure 3 come out of the
inner regions of the accretion flow. In \citet{pun07} and in Chapter
11 of \citet{pun01} it was shown that the Poynting jet that is
driven from the inner edge of the accretion is tantamount to the
high accretion rate analog of the "ergospheric disk" jet that was first
described in \citet{pun90}.
\begin{figure}

\includegraphics[width=100 mm]{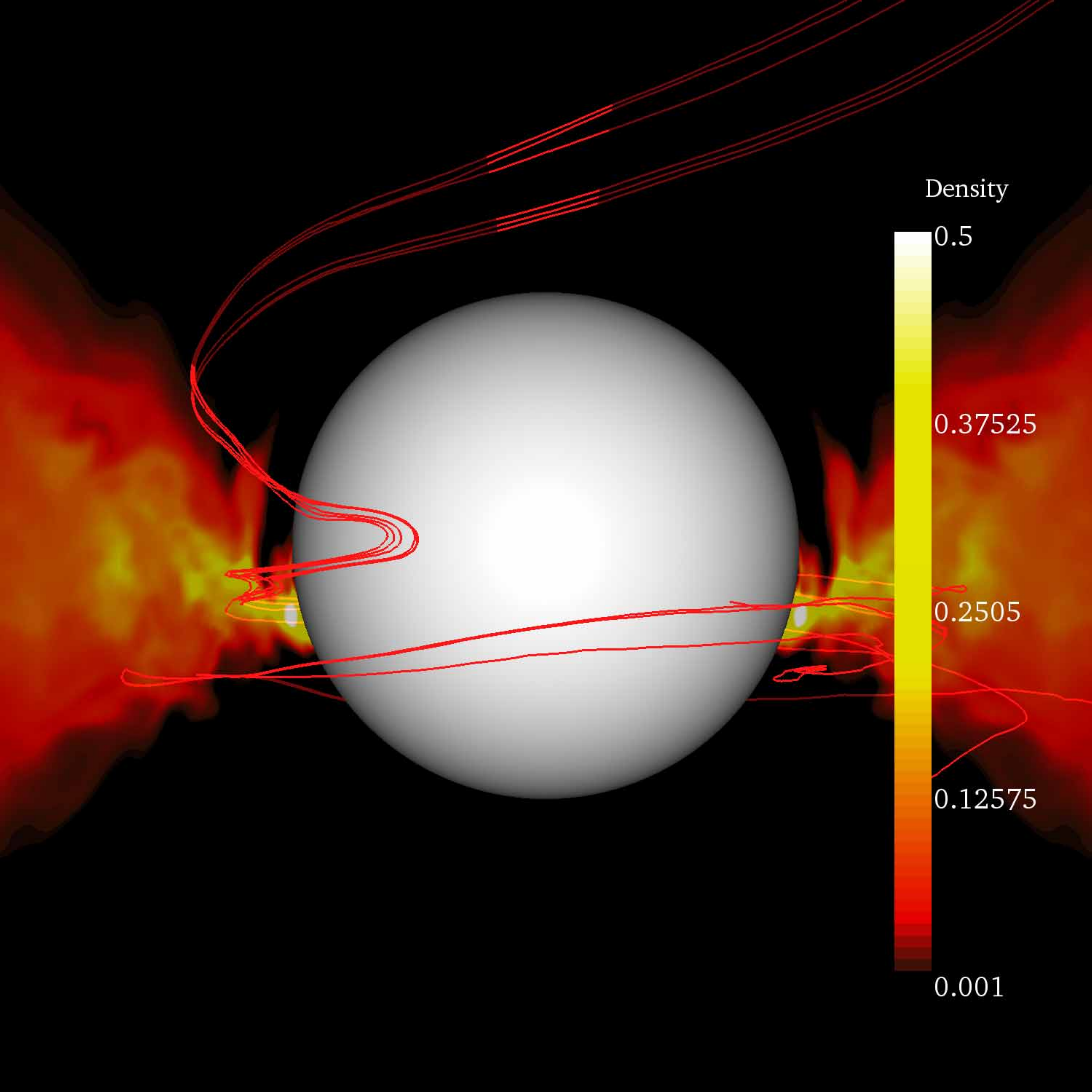}

 \caption{The type I vertical magnetic field lines that emerge from the inner equatorial accretion flow. The false color plot is
 a contour map of a 2-D cross-section of the density in Boyer-Lindquist coordinates expressed in code units.
 The interior of the inner calculation boundary is grayish-white. These field lines are distinguished by connecting
to the Poynting jet in one hemisphere only, with the other end
spiraling around within the accreting gas in the opposite
hemisphere. Another distinguishing feature of the type I field lines is that the azimuthal
direction of the magnetic field changes
direction as the field line crosses the midplane of the accretion
flow.}
\end{figure}
\begin{figure}

\includegraphics[width=100 mm]{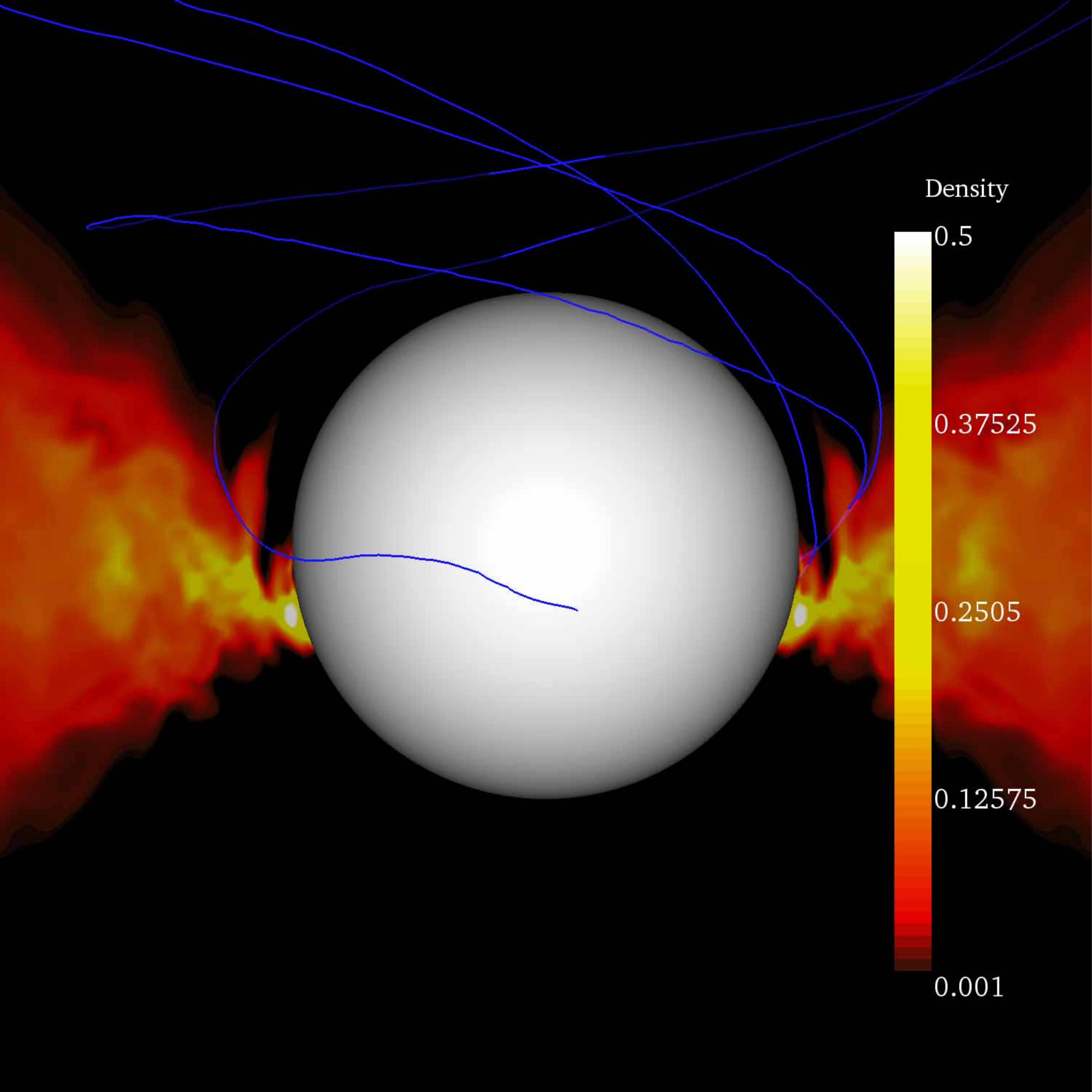}

 \caption{The type II vertical magnetic field lines that emerge from the inner equatorial accretion flow. The false color plot is
 a contour map of a 2-D cross-section of the density in Boyer-Lindquist
 coordinates expressed in code units. The interior of the inner calculation boundary is grayish-white. The type II field lines
 intersect the inner calculational boundary. They traverses the lateral
extent of the accretion flow, connecting the jet in one hemisphere
to their point of origin on the inner calculational boundary in the
opposite hemisphere. Notice that the type II field lines interact
mainly with the less dense regions (typically, 1\% - 10\% of the
peak density) of the accretion flow.}
\end{figure}
\begin{figure}

\includegraphics[width=100 mm]{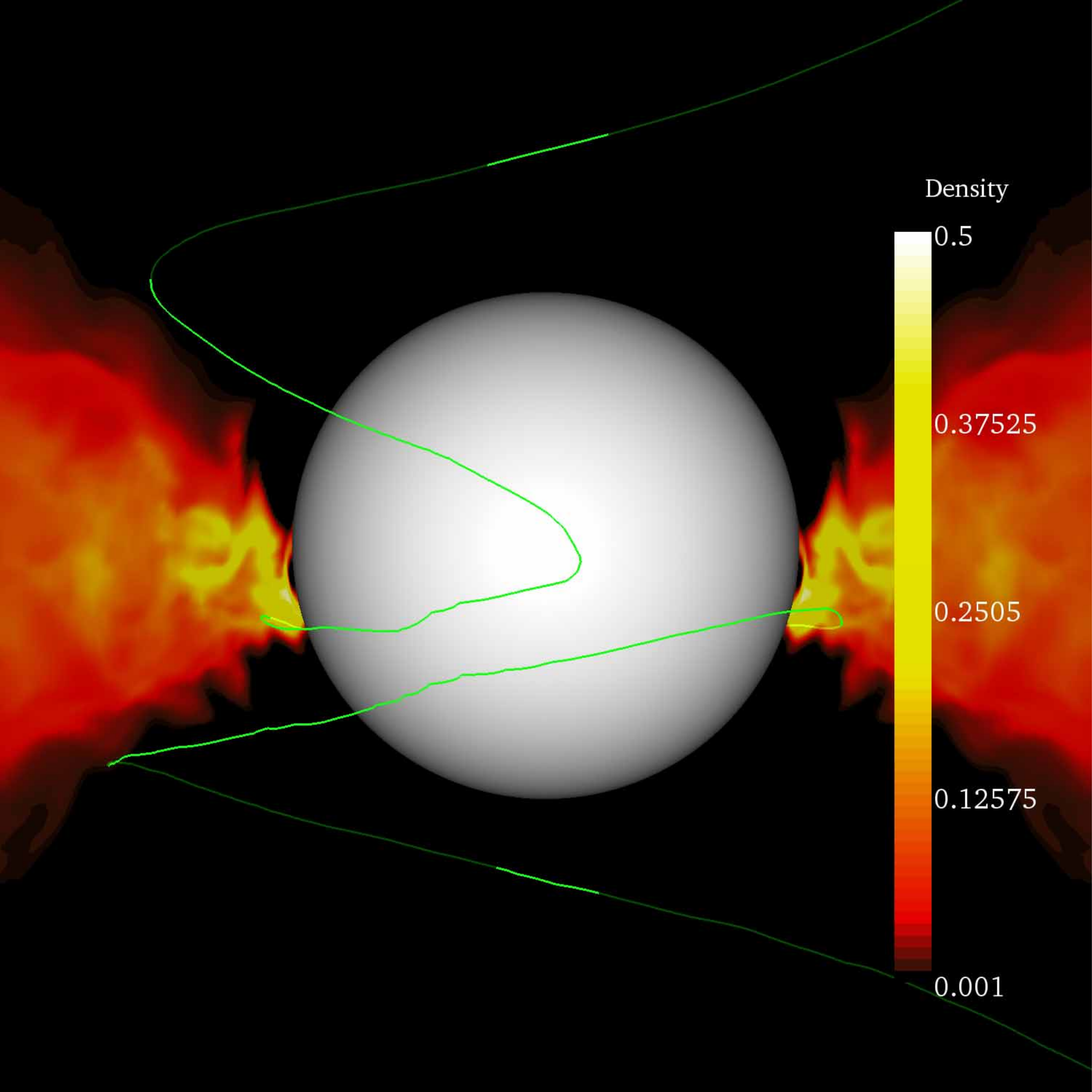}

 \caption{The type III vertical magnetic field lines emerges from the inner equatorial accretion flow. The false color plot is
 a contour map of a 2-D cross-section of the density in Boyer-Lindquist coordinates
 expressed in code units. The interior of the inner calculation boundary is grayish-white. The type III field lines
connect to both sides of the bipolar jet. Like the type I field lines the azimuthal
direction of the magnetic field changes direction as the field line crosses the midplane of the accretion
flow.}
\end{figure}
\begin{figure}
\includegraphics[width=110 mm, angle=0]{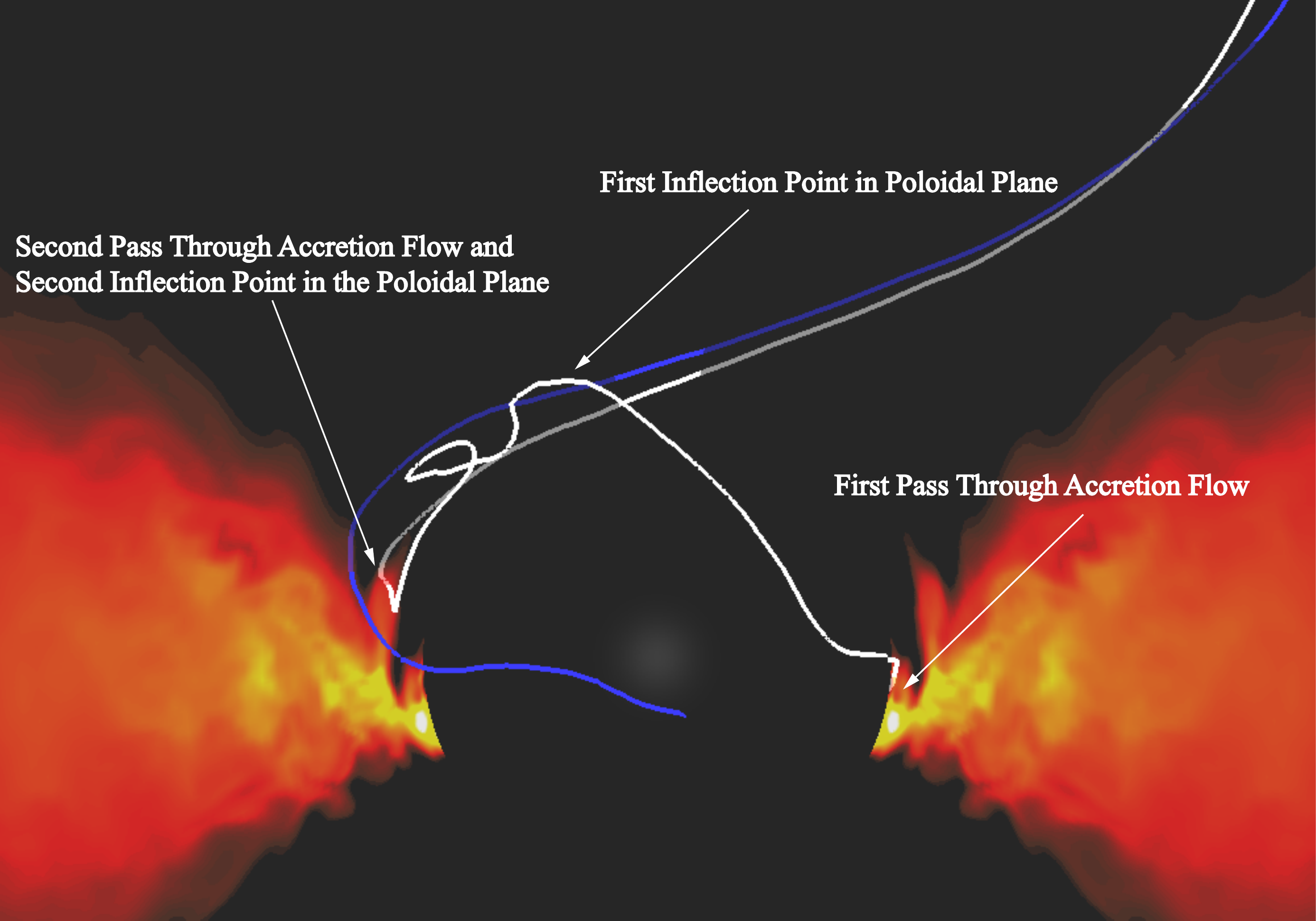}

 \caption{A type IV field is shown in white. The false color plot is
 a contour map of a 2-D cross-section of the density in Boyer-Lindquist coordinates. The interior of the inner calculation boundary is black.
 This figure compares and contrasts the type IV and type II field lines (shown in blue for comparison).
 These field lines are from figures 4 and 5, but they have been individually rotated in azimuth by different amounts for illustrative purposes.
 Thus, there is no physical connection between the density contours and the individual field lines. Both field lines intersect
 the inner calculational boundary. The type IV field line intersects the boundary at a slightly higher latitude than the type II field line.
 The initial integration away from the
 inner calculational boundary proceeds similarly, both field lines increase in latitude as they spiral around the black hole, exiting the accreting plasma.
 The type II field line is at slightly larger radius and connects directly into the jet once it exits the accreting plasma. By contrast,
 the type IV field line experiences an inflection in the poloidal plane and bends back down towards the accretion flow.
 The field line continues its spiral around the black hole dipping into the upper layers of the accretion flow one more time, at a larger radius than
 the first pass through the accretion flow. Within these upper layers it experiences a second poloidal inflection point and bends upward,
 finally merging with the Poynting jet.}
\end{figure}
\subsubsection{Vertical Field Line Taxonomy}
\par We now describe the different types of vertical field lines that
were found and motivate the distinct designations that were assigned
by defining characteristics that distinguish each type. In all of
the figures that display the magnetic field lines in KDJ in this
paper, the poloidal field goes from the northern hemisphere towards
the southern hemisphere (correcting the mistake that the field lines
point from south to north in \citet{pun07,pun08}). Consequently, we
did not include arrows on the field lines because they added
additional clutter to the already complex, very inhomogeneous
environment. The vertical field line categories are labeled type I -
type IV. The type I - type III field lines are illustrated
explicitly in figures 6 - 8, respectively.
 \subsubsubsection{Type I Field Lines} These field lines are distinguished by connecting to
the Poynting jet in one hemisphere only, with the other end
spiraling around within the accreting gas in the opposite hemisphere
(see figure 6). Another distinguishing feature of the type I field
lines is that the azimuthal direction of the magnetic field changes
direction as the field line crosses the midplane of the accretion
flow. Figure 6 is viewed along the equatorial plane with the
Paraview camera positioned at an azimuth similar to its location in
figure 5. Half of the jet field lines from the inner accretion flow
at t = 9840 M (11 out of 22) are this type in figures 4 and 5 (note
that the frequency of occurrence of a type of field line does not
equate to a magnetic field strength). In this time step, type I
field lines are the most common variety in the ergospheric disk jet.
They are color coded in red in the figures throughout this article.
The "anchoring" end of the field line spirals just below the equator
near the black hole for many revolutions, then spirals out into the
distant reaches of the corona or accretion disk proper within the
hemisphere opposite the jet. Thus, the anchoring end of the field
line becomes inextricably tangled with the other toroidal coils
within the turbulent region of the accretion flow. This field line
topology was never envisioned in theoretical work. Dynamically speaking,
the most important feature of the type
I field lines in figure 6 is that the azimuthal direction of the
magnetic field changes direction as the field line crosses the
midplane of the accretion flow. This type of magnetic field
polarization reversal is associated with an Alfven mode in MHD and
is sometimes called an Alfven rotation \citep{kap66}. The type I
field lines are a conduit for powerful Alfven waves to radiate from
the accretion flow into the jet within the accretion vortex.
\subsubsubsection{Type II Field Lines} The type II field lines
originate from the inner calculational boundary. They traverses the
lateral extent of the accretion flow, connecting the jet in one
hemisphere to their point of origin on the inner calculational
boundary in the opposite hemisphere. This class of vertical field
line was never anticipated theoretically. They are depicted with the
examples extracted from time step t = 9840 M in figure 7. This is
the second most common type of field line in the ergospheric disk
jet at t = 9840 M in figures 4 and 5 (6 out of the total of 22 field
lines). However, considering all three time snapshots to be
discussed, it is the most common, 29 out of the total of 63
ergospheric disk jet field lines that are plotted in this section
(see the discussions in subsections 4.3 and 4.4). The false color
contour plots of the density in figures 4, 5 and 7 indicate that the
type II field lines (color coded blue) interact mainly with the less
dense regions (typically, 1\% - 10\% of the peak density) of the
accretion flow. The interaction of the field lines with the
accreting gas creates outgoing Poynting flux that is transmitted
along the poloidal field lines from the accreting gas into the jet.
\subsubsubsection{Type III Field Lines} The type III field lines
(color coded green) connect to both halves of the bipolar jet (see
figure 8). The type III field line that is illustrated in figure 8
was extracted from the jet in figures 4 and 5. This was the
type of vertical field line that had been expected to occur in
equatorial accretion flows near black holes in simplified
theoretical discussions \citep{thp86,pun90,pun01}. Surprisingly,
this turns out to be the rarest type of field line that we
encountered in KDJ. Only 2 out of 22 jet field lines were of this
type at t = 9840 and only 6 out 63 in all three time slices
combined. Like the type I field lines, the azimuthal field direction
switches sign within the accretion flow near the equatorial plane
(the Alfven mode polarization of MHD). The scarcity of type III
field lines is intimately related to the fact that the ergospheric
disk jets form asymmetrically in KDJ. In all three time slices, one
jet (at t = 9840 M it is clearly the northern jet) is much more
powerful than the other. We have color coded the type III field
lines as green throughout. We note that the 3-D numerical
simulations of the relativistic string approximation to thin
magnetic flux tubes around rapidly rotating black holes in
\citet{sem04} produced type III field lines exclusively.
\par In figure 8, the Paraview camera is rotated $70^{\circ}$ about
the black hole rotation axis relative to figures 5-7. Notice that
the gas density in the inner regions of the cross-section is
considerably larger than in figures 6 and 7. The accretion flow is
clearly nonuniform in azimuth. Similar dense "fingers" of accreting
gas also occurred in the 3-D numerical simulations of \citet{igu08}
as exemplified by the even columns of figure 2. This appears to be
an endemic feature of 3-D accretion in the presence of vertical
flux, since similar "tongues" of dense plasma occur in 3-D MHD
simulations of gas accreting onto highly magnetized stars
\citep{rom08}.

\subsubsubsection{Type IV Field Lines}The type IV field lines follow
a circuitous path from the inner calculational boundary to the jet.
Ostensibly, the integral curves near the inner calculational
boundary seem to be on a trajectory that will either promptly merge
with the jet after passing through some accreting gas (just like the
type II field lines) or they might seem to be destined to link the
horizon to the jet directly (like a Blandford-Znajek type field
line). However, instead of a direct path to the jet, the type IV
field lines are diverted back towards the accretion flow (This is
explicitly demonstrated by figure 9 which contrasts a type IV field
line with a type II field line). The type IV field lines are
characterized by this inflection point in the poloidal plane. These
field lines bend back towards the accretion flow, penetrating the
surface layer of the inner edge of the accretion flow, then
"reflect" out of the accreting flow into the jet. The inner portion
of the field line is effectively one half of a magnetic field loop,
either connecting the innermost accretion flow to a more distant
part of the accretion flow, or connecting the inner calculational
boundary to the accretion flow. But instead of the magnetic loops
closing, the field lines are redirected back out of the accretion
flow into the jet - in the same hemisphere that contains the
"half-loop". Thus they have two inflection points in the poloidal
plane. The first bends the field line from above the accretion flow
down into the accretion flow. The second inflection point bends
bends the field line out of the accretion flow into the Poynting
jet. The multiple inflection points make these the most complicated
field lines that are considered. The type IV field lines (color
coded white) in figures 4 and 5, propagate outward from the inner
calculational boundary, but are corralled by a wall of type II field
lines. Visually, it appears that this is related to the inflection
point that bends the type IV field toward the accretion flow. The
type IV field lines gradually (sometimes after more than one revolution
about the black hole) work their way through gaps in the type II
magnetic wall at t=9840 M. The multiple inflection points and the
fact that they traverse one or more complete circuits about the
black hole, make the type IV field lines the most difficult to
visualize without a fluid camera angle.
\begin{figure}

\includegraphics[width=100 mm]{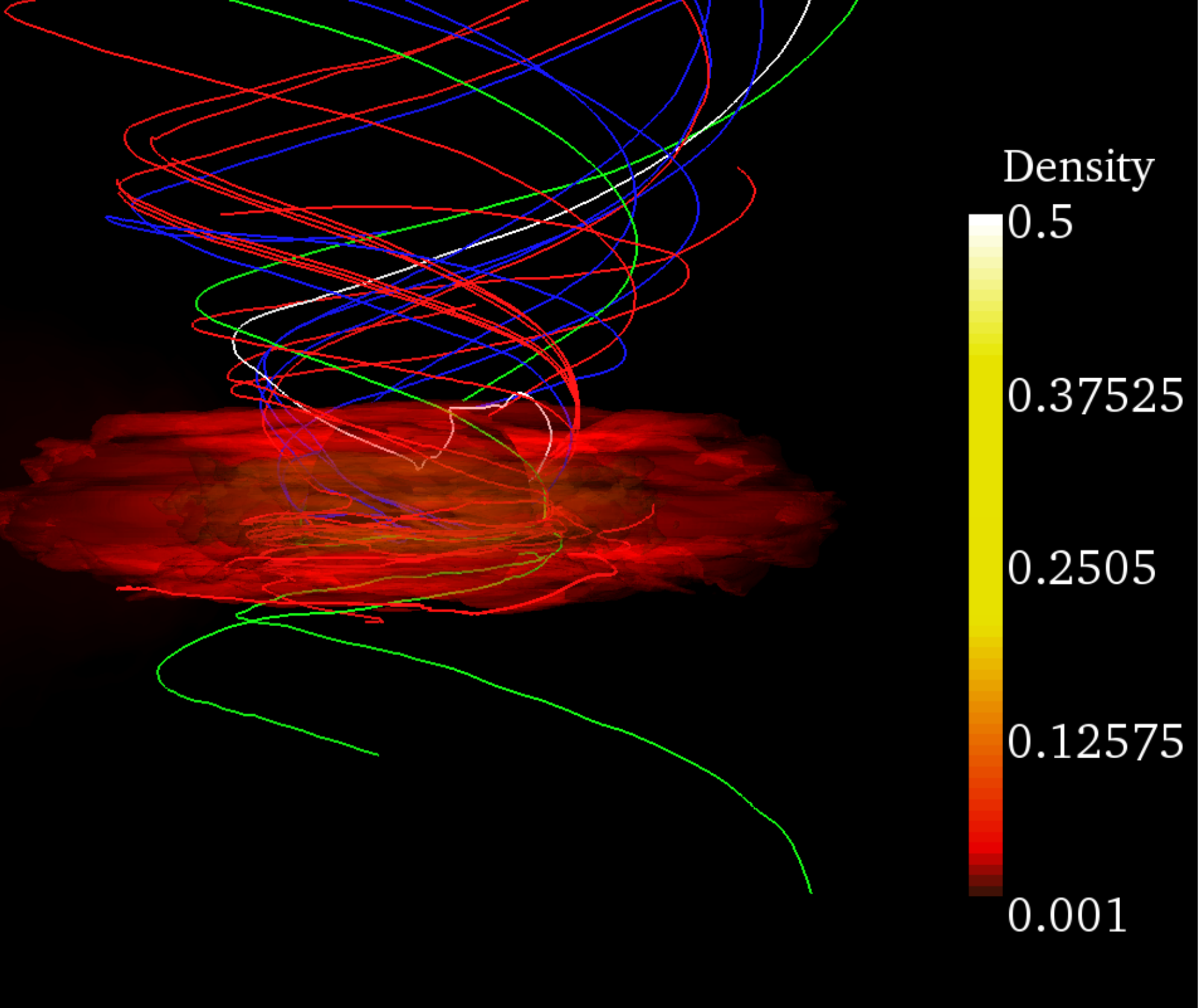}
\caption{The first frame of movie 2. The false color background is
comprised of three 3-D contours of the Boyer-Lindquist density
superimposed on each other. One is at 0.05 in code units and is
plotted with 40\% opacity, the next contour level is at 0.1 with
35\% opacity and the third contour level is at 0.15 and it is
plotted at 30\% opacity. Notice that the white, type IV field line
emerges from accreting gas on the right then bends down back into
the accretion flow briefly in the front before joining the jet. This
second encounter with the accretion flow is the defining
characteristic of type IV jet field lines.}
\end{figure}
\subsubsection{Animations Reveal a Complicated Layered Magnetosphere}
\par Every line of sight has hidden features in 3-D. Clearly there are
dynamically important features in figure 5 within the accretion flow
that can not be seen in figure 4. Animations of the camera rotating
around the black hole are the best way to visualize these complex
3-D magnetospheres. Thus, we have supplied an animation of the
camera rotating counter-clockwise about the black hole spin axis
(the vertical axis in figures 4 and 5) with the line of sight
restricted to the equatorial plane (as in figures 4 and 5),
designated as movie 2 in the on-line material. The first frame of
this animation is given in figure 10. In order to keep the gas
semi-transparent, we have plotted the density a bit differently in
the animation than was done in figures 4 and 5. For the sake of
maximum clarity, we have plotted three, 3-D, contours of the
Boyer-Lindquist density superimposed on each other. One contour
level is at 0.05 in code units and is plotted with 40\% opacity, the
next is at 0.1 with 35\% opacity and the third contour level is at
0.15 and it is plotted at 30\% opacity. Three, 3-D contours surfaces
and this level of transparency gave the best balance between seeing
the gas and seeing field lines within the gas in our opinion.  We
only plot one of the type IV field lines because a plot of all 4
such field lines circling the black hole blocks our view of the
other field lines in the jet. Unfortunately, the transparency of the
gas in the animation is not readily apparent in the print version of
the first frame (figure 10). The back illumination of the images
that are provided by a computer screen enhances the transparency of
the translucent gas contours.
\begin{figure}
\includegraphics[width=150 mm]{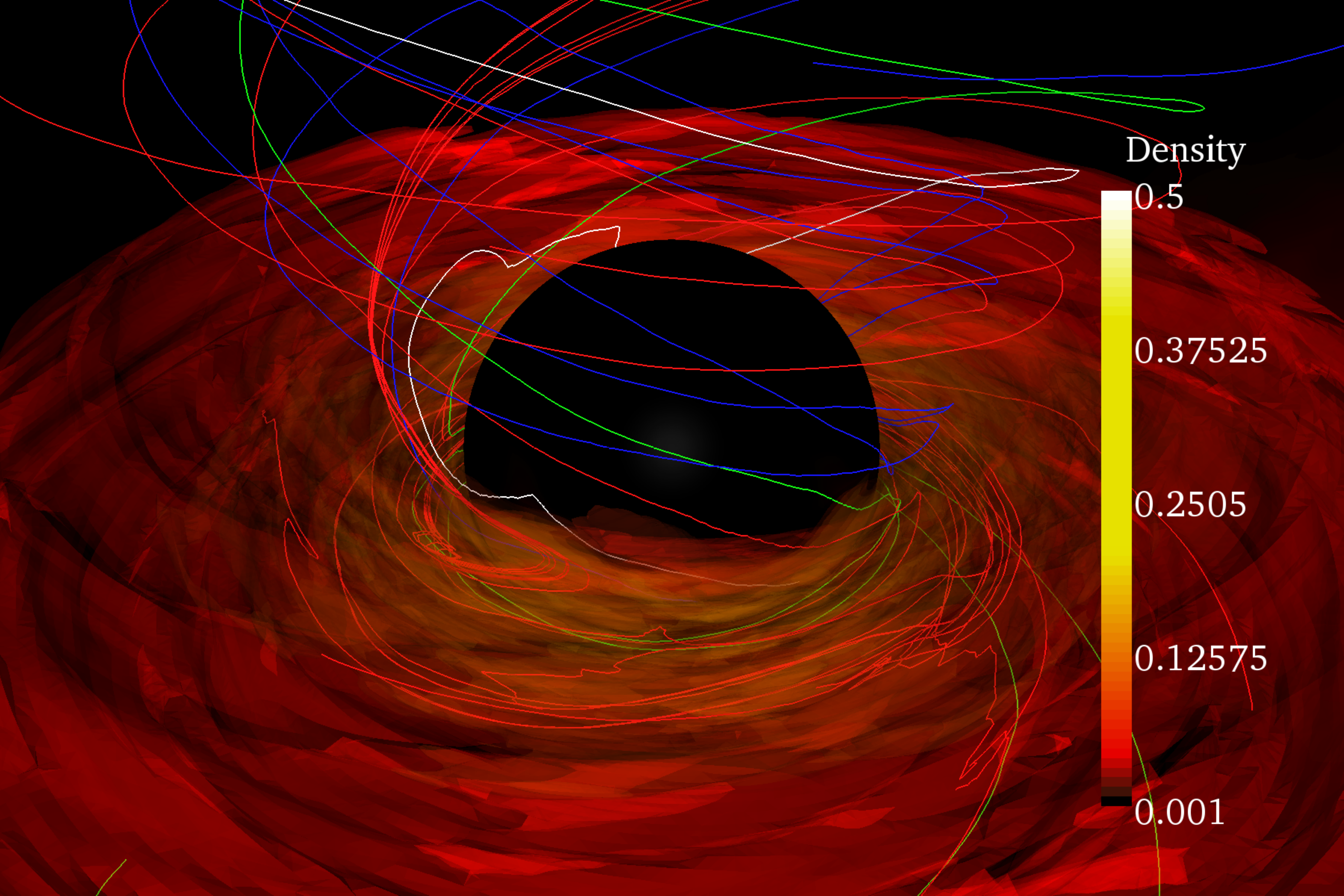}
 \caption{The first
frame of "isometric view" movie 3. The false color background is
comprised of three 3-D contours of the Boyer-Lindquist density
superimposed on each other. One is at 0.05 in code units and is
plotted with 40\% opacity, the next contour level is at 0.1 with
35\% opacity and the third contour level is at 0.15 and it is
plotted at 30\% opacity.}
\end{figure}
\par Movie 2 and figure 10 are useful for seeing the jet emerging from
the accreting gas. However, this equatorial line of sight obscures
the interplay between the base of the jet field lines and the
accreting gas. The maximum amount of clarity is given by the
"isometric view" (a line of sight that is 30 degrees above the
equatorial plane). This view allows for a clear line of sight
towards the interaction of the base of the field lines that comprise
the jet with the accreting gas (see movie 3 and figure 11). The jet
can still be seen, but many of the coiled jet field lines attain
peculiar shapes due to the off-angle perspective. From this angle,
it would be hard to realize that the northern jet is conical as is
obvious in movie 2 (figure 10). Figure 11 is the first frame of
movie 3, the isometric view. In movie 3, the Paraview camera swings
around the black hole rotation axis in the counter-clockwise
direction, all the time maintaining a latitude of $30^{\circ}$ above
the equator. The false color background is comprised of three 3-D
contours of the Boyer-Lindquist density superimposed on each other
at the same levels and opacity as figure 10 (movie 2). Again, for
the sake of clarity, we plot only one of the type IV field lines.
The value of the gas density in regions of the accretion flow that
are threaded by jet field lines is more readily apparent in movie 3
than in figure 11. This is an example of how a fluid camera angle is
vastly superior to simple snapshots and slices for clarifying
complicated structures. As the camera angle starts to rotate in
movie 3, one can see that a cluster of type I field lines, just to
the left of center. At this camera angle and in figure 11, it
appears as if the cluster of type I field lines is embedded within a
"finger" of the higher density gas. However, during the time
interval 5 to 7 seconds in movie 3, it is revealed that the field
lines actually reside in a low density cavity that is adjacent to
this "finger" of dense gas. From the isometric perspective in figure
11 and movie 3, it is clear that all of the jet field lines emerge
from the accreting gas and do not go straight from the event horizon
into the jet as proposed in the model of \citet{blz77}. The
existence of a significant number of these jet field lines (that do
not go directly from the horizon to jet) is in contradistinction to
the findings of \citet{hir04} that such field lines do not exist in
the particular 3-D simulation that they studied in detail.

\par Our detailed study of the field line types in the animations indicate
the basic following structure of the inner disk magnetosphere at t =
9840 M. The inner-most jet field lines from the ergospheric
accretion flow are type IV. Just outside of these, the type II field
lines appear to thread the gas in an irregularly shaped inner edge
(or boundary layer) of the accretion flow that angles across the
equatorial plane. In figures 4, 5 and 7, it is clear that the type
II lines intersect the accretion flow in regions on the north face
where the density is only about 1\% - 10\% of the density peak. The
strongest azimuthally averaged poloidal field strength in this time
slice is within the base of the jet emerging from the ergospheric
accretion flow. The magnetic pressure of these field lines is so
large that it is apparently redirecting the accreting gas to flow
predominantly below the equatorial plane as it approaches the inner
calculational boundary. In figure 11 (and movie 3), just outside of
the type II lines are the type I lines. As discussed above, in figure 11 (and movie 3),
we can see that the type I field lines carve out low density
cavities ($\sim$ 10\% of the peak density) within the accreting gas
above the midplane. Finally, the type III field lines seem to be
located the farthest out in the accretion flow at t = 9840 M, but
this conclusion is based on only two type III field lines.
\subsubsection{Magnetically Arrested Accretion in 3-D}
\par There is clearly a strong interaction between the magnetic
pressure from the vertical flux forming the base of the jet and the
accreting gas. We explore this interface and its relationship to
magnetically arrested accretion that was discussed in the context of
simulations in the Pseudo-Newtonian potential in section 3. Figure
12 is an ambitious attempt to capture three dynamical entities in
one figure, the "streamlines" of the accreting gas, the magnetic
field lines and the density. Unfortunately, the complicated 3-D dynamics
requires a complex figure and Paraview 3.3.0 gives us the visualization tool
to create relatively easy to understand images. In figure 12, the Boyer-Lindquist
evaluated gas density is seen in cross-section by means of a false
color contour map. The color bar on the right hand side of the
figure is in code units and the scale was optimized to give the best
contrast between the field lines, the accretion flow gas density and
the background. The magnetic field lines were chosen to lie within
the magnetic wall located near the inner edge of the northern side
of the accretion disk. They were preferentially selected to
intersect the gas in the plane of the image. The magnetic wall that
is referred to here is the same as the magnetic wall that was
discussed in subsection 4.2.2. This choice of plotting preferences
has the tremendous advantage that one can see the density of the gas
in which the magnetic wall is embedded. In order to elucidate the
structure of the magnetic wall, the opacity of the gas was chosen to
be relatively high (85\%), so that it was obvious when the field
lines penetrated the gas within the cross-sectional slice. In
particular, the field lines comprising the magnetic wall on the left
hand side of figure 12, suddenly almost completely disappear when
they intersect the red false color gas density contour. On the left
hand side, we see a strong magnetic wall of type II field lines. On
the right hand side, the magnetic wall appears to be a roughly equal
mix of type I and type II field lines. They overlay almost on top of
each other, so the red and blue meld into a maroon color.
\begin{figure}
\includegraphics[width=150 mm]{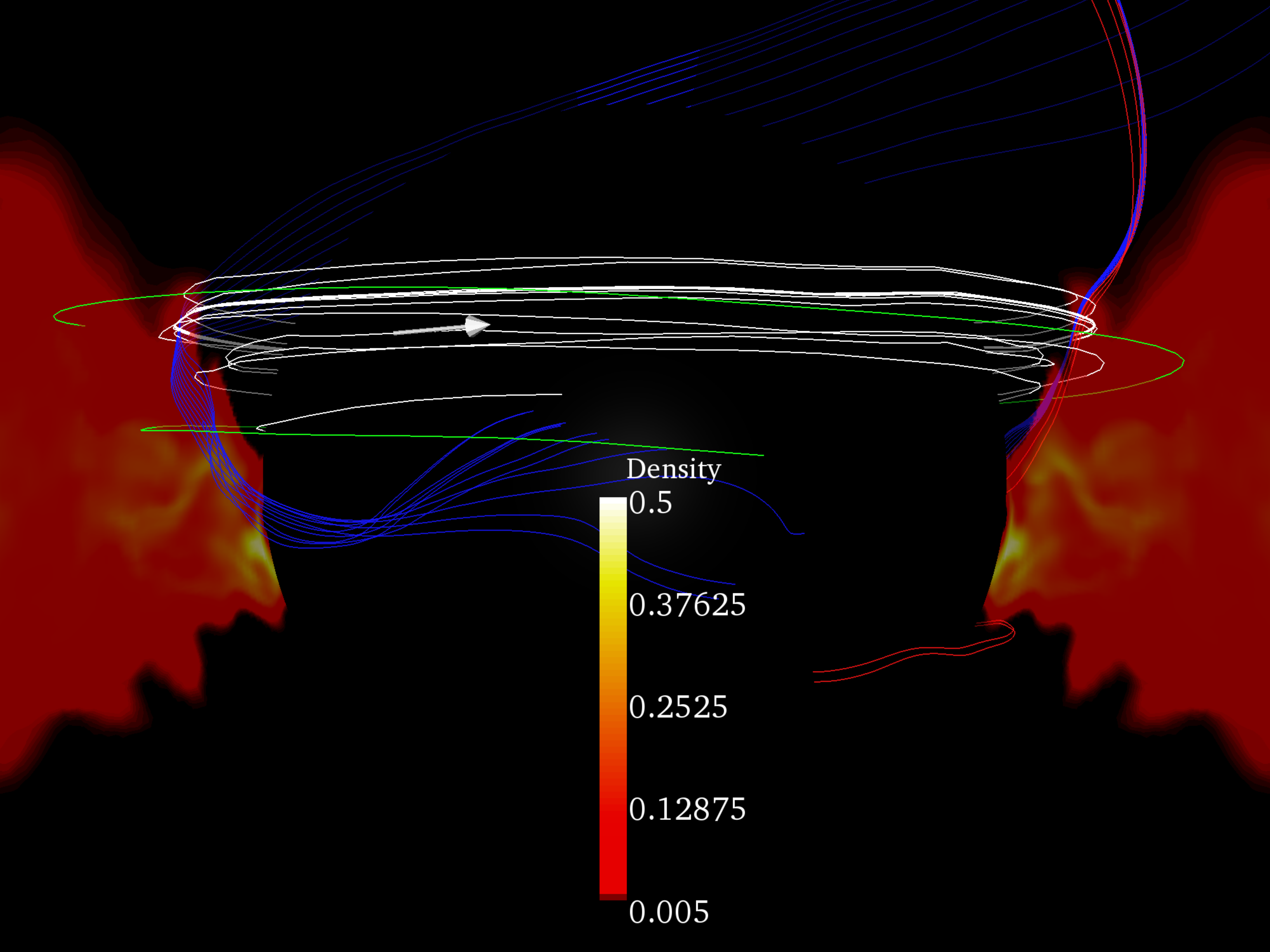}
\caption{The 3-D version of magnetically arrested accretion. The
white and green curves are integral curves of the Boyer-Lindquist
evaluated velocity at t = 9840 M, gas streamlines. The false color
contour map of a cross -section of the density is in code units. The
blue and red curves are magnetic field lines with the color coding
described in the text. The white curve is a magnetically arrested
streamline. Notice that as the footpoint of the type II field lines
on the inner calculational is displaced toward the north, the topology
starts to deform toward that of the type IV field lines.}
\end{figure}
\par On the backdrop of the magnetic wall and the accretion flow, we
plotted two streamlines of the accreting gas defined as the integral
curves of the velocity (the analogs of equations A.16 - A.18),
\begin{eqnarray}
&& \frac{dx}{ds} = \frac{(v^{r})\sin\theta\cos\phi + (rv^{\theta})\cos\theta\cos\phi - (r\sin\theta v^{\phi})\sin\phi}{\|v\|} \\
&& \frac{dy}{ds} = \frac{(v^{r})\sin\theta\sin\phi + (rv^{\theta})\cos\theta\sin\phi + (r\sin\theta v^{\phi})\cos\phi}{\|v\|} \\
&&
\frac{dz}{ds}=\frac{(v^{r})\cos\theta-(rv^{\theta})\sin\theta}{\|v\|} \\
&& \|v\|^{2} \equiv g_{rr}{v^{r}}^{2}+
g_{\theta\theta}{v^{\theta}}^{2}+g_{\phi\phi}{v^{\phi}}^{2}\;,
\end{eqnarray}
where, $v^{r} = dr/dt$, $v^{\theta} = d\theta / dt$ and $v^{\phi}=
d\phi / dt$. It is important to clarify for the reader that these
are not true streamlines in the time dependent sense. The green and
white lines do not represent global trajectories into the black
hole, since the vertical magnetic field is dynamic and therefore so
are the magnetic forces in KDJ; the magnetic forces could be
significantly altered after a few revolutions of the black hole. For
this reason, the trajectories are only physically representative of
the gas propagation in a local sense. We chose two starting points
for the integration of the gas streamlines for illustrative
purposes. The first integration resulted in the white integral curve
that was selected because it was strongly perturbed from rapid
radial in-fall at the location of the magnetic wall on the left hand
side. We added only one arrow (to avoid clutter) to show the
direction of motion in both streamlines. The white curve shows us
two things. Notice that the integral curve seems to be closely
aligned with the kinks seen in the magnetic wall. These kinks could
represent local enhanced magnetic stresses that naturally arise when
gravitational forces on the gas and pressure gradients within the
gas are in opposing directions. Also notice that the accretion of
the plasma that is represented by this streamline is clearly
magnetically arrested. It takes many spirals around the black hole
before it crosses the magnetic barrier by winding through small gaps
and finally reaches the inner calculational boundary. To show that
this streamline actually represents magnetically arrested gas, we
plotted a "control" streamline in green. This streamline is chosen
at a similar disk height, but farther out in the disk. It never
strongly interacts with the magnetic wall. It reaches the inner
calculational boundary at an azimuthal coordinate just beyond the
"footpoints" of the magnetic wall, flowing always outside the
magnetic wall. Thus, it flows toward the inner calculational
boundary virtually unimpeded by any of the putative strong magnetic
forces associated with the magnetic wall. The green streamline in
figure 12 is typical of most streamlines in the body of the
accretion disk and indicates that a direct inflow takes only
$\approx 1.5$ spirals to reach the inner calculational boundary, a
much shorter journey than the white (magnetically arrested)
streamline.
\par Even
though the white streamline does not represent a global time
dependent path for the gas (because of the vertical field and
associated magnetic force variations noted above), it is indicative
of a strong interaction all the along length of a slow inward
spiral. The magnetic kinks and the low pitch angle spiral of the
streamline are consistent with the notion that the gas is being held
up by magnetic forces virtually everywhere all along the length of
the streamline at the interface of the magnetic wall. The
magnetically arrested property is depicted by figure 12, but the
number of spirals around the black hole that it takes for a given
fluid element to reach the black hole is not determined by this
integration. Even though some gas is magnetically arrested, the
majority of the accreting gas finds its way to the black hole by
maneuvering around the strong patches of vertical flux, passing
through the azimuthal gaps in the vertical magnetic flux
distribution and the vertical gaps in the azimuthal twists in the
type I field lines near the equatorial plane.

\subsection{The time-slice t = 9920 M} The next time step in which
data was dumped is at t = 9920 M. Unfortunately, this time slice is
too distant in time from the t = 9840 M time slice data in order to
define any continuous change in magnetospheric parameters ($\gtrsim
6$ black hole rotational periods). The most surprising element of
this time slice is that the stronger jet in figure 13 points to the
south, the opposite of the case for t = 9840 M. The magnetospheric
geometry at t = 9920 M is the least complex of any that is found in
the three available time slices. Thus we only use 40 starting points
for our field line integrations. The starting points of integration
are restricted to $\pm 4^{\circ}$ of the equatorial plane as before.
With this restriction on latitude, we chose 40 points at random from
the interval, $1.30 M < r < 1.60 M $.
\begin{figure}
\includegraphics[width=150 mm]{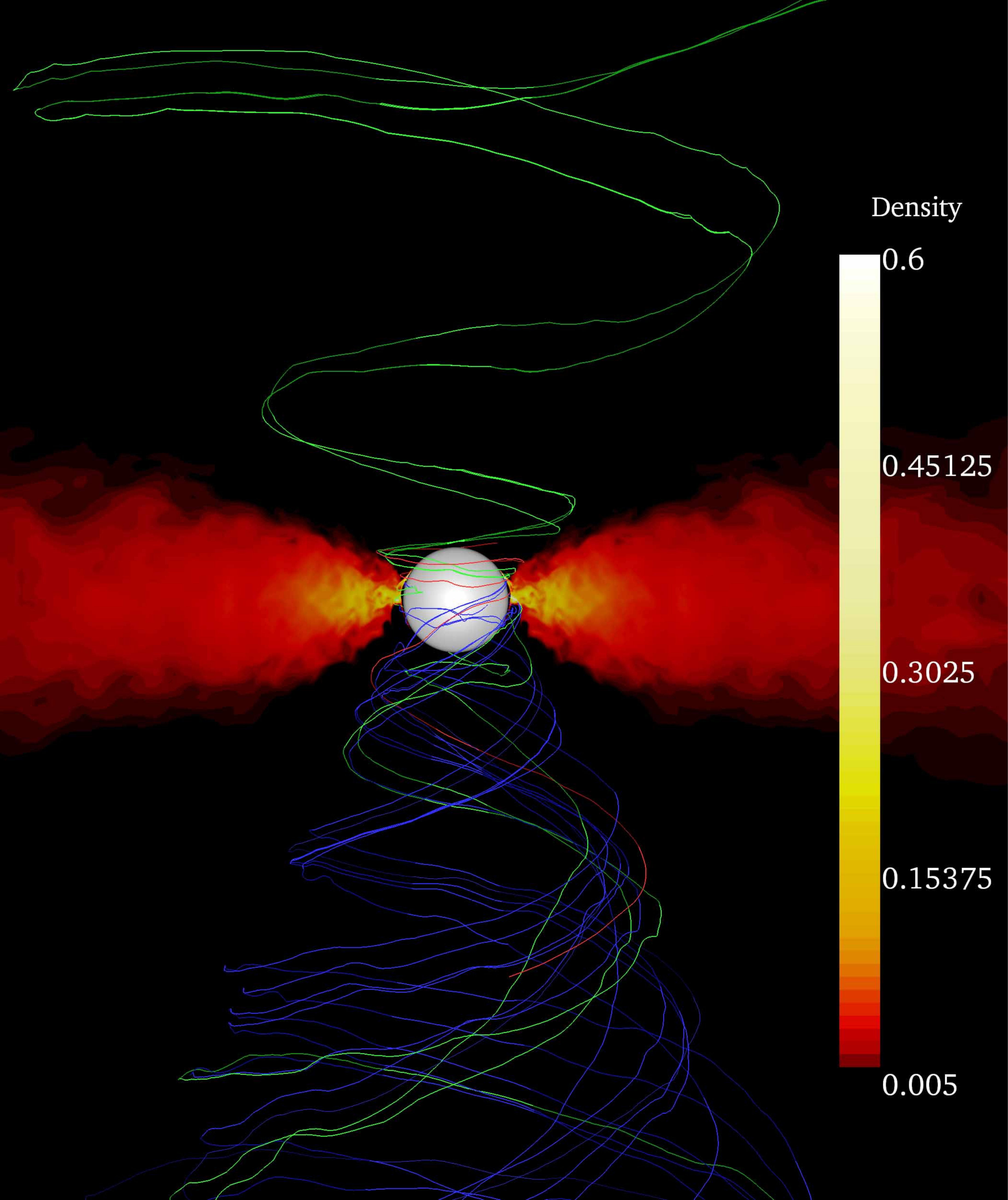}
\caption{The wide angle view of the ergospheric disk jet at t = 9920
M. This is a false color plot of a cross-section of the
Boyer-Lindquist evaluated density. The color bar on the right hand
side is in code units. The magnetic field lines are color coded as
described in section 4.3}
\end{figure}
\par Figure 13 is a false color contour plot of a cross-section of the
Boyer-Lindquist density in code units. Of the total of 40 field
lines that were plotted, 15 were jet field lines. In this time slice
there were 1 type I field line, 11 type II field lines, 3 type III
field lines and 0 type IV field lines. So if the time slice, t =
9840 M, could be construed as a predominantly a type I event to the
north, the time slice, t =9920 M, could be considered a type II
event to the south. The field lines emerging to the south from the
inner edge of the accretion flow explain the direction of the white
arrows (azimuthally averaged poloidal Poynting vector) in the top
right hand frame of figure 3. These field lines are the source of
the powerful narrow channel of radial Poynting flux that emerges
from the ergosphere in the top right frame of figure 3. Notice that
the field lines that form the southern jet displace large volumes of
gas as they exit the inner accretion flow. Low density, magnetic
regions (islands) are created just below the equatorial plane at the
base of the jet. We note that the significant number of type III
field lines in the northern jet could not have been anticipated from
the azimuthally averaged plot in the top right hand frame of figure
3. Careful inspection of figure 13 shows that two of the type III
field lines lie almost on top of one another in the northern jet.
\subsection{The time-slice t = 10000 M}
The magnetic field configuration at t = 10000 M, near the black
hole, is the most complicated of the three time slices. There is a
strong jet to the north that is very similar to the southern jet
that was seen at t = 9920M and there is a modest southern jet. There
was evidence of this southern jet in the azimuthal average of
$S^{r}$ in the bottom frame of figure 3 in the form of two
disconnected patches of red and yellow. However, it was unclear if
this represented two pieces of a continuous jet and there was not
enough visual evidence to determine the 3-D magnetic field
configuration that was responsible for these regions of enhanced
$S^{r}$. Figure 14 is an alternative visualization to the wide angle
view of a jet that was presented in figures 4 and 13. Instead of a
false color contour plot of a cross-section of the density, we plot
two 3-D density contours overlayed on top of each other, similar to
figure 10. Such a representation of the density is conducive to a
3-D visualization of the jet/accretion flow interaction (since the
gas is semi-transparent for all lines of sight) that is optimally
displayed in the form of an animation of the Paraview camera
rotating through different viewing angles to the black hole. Figure
14 is the last frame of the on-line animation, movie 4. The density
contours are chosen at 0.05 and 0.1, with opacities of 40\% and
35\%, respectively. The line of sight from the Paraview camera to
the black hole is in the equatorial plane. The camera swings around
the black hole spin axis in the counter-clockwise direction.
\begin{figure}
\includegraphics[width=160 mm]{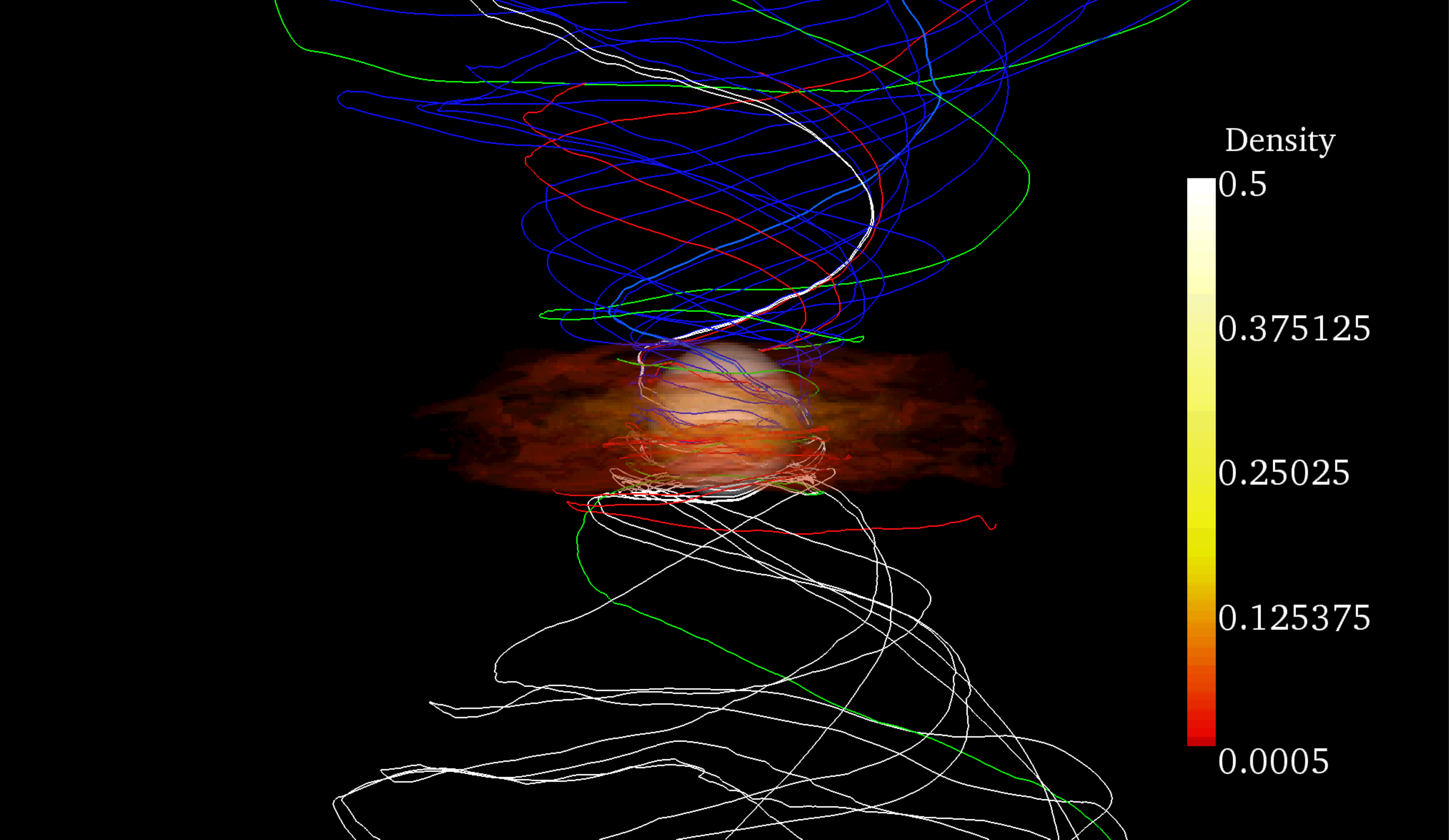}
\caption{The wide angle view of the ergospheric disk jet at t =
10000 M. The image is the last frame of the on-line animation, movie
4. The density contours are chosen at 0.05 and 0.1, with opacities
of 40\% and 35\%, respectively. The magnetic field lines are color
coded as described in section 4.3. The interior of the inner
calculation boundary is grayish-white.}
\end{figure}
\par Figure 14 and movie 4 show bipolar jets propagating away from the
inner accretion flow. The latitude of the starting points of
integration was restricted to $\pm 4^{\circ}$ of the equatorial
plane. With this restriction on latitude, we chose 20 points at
random from the interval, $1.35M < r < 1.55 M$, and an additional 40
starting points for integration were chosen randomly from the
interval, $1.30 M < r < 1.60 M $. This is the opposite of what was
chosen at t = 9840 M. The motivation for this change is that the
bottom frame of figure 3, indicated that there was likely some
interaction of the field lines comprising the southern jet with the
accreting gas near $r \lesssim 1.6 M$. A total of 27 out of the 60
field lines connect to the Poynting jets, 18 field lines go to the
northern jet and 9 to the southern jet. Out of the 18 field lines
that comprise the northern jet, 3 are type I, 12 are type II, 1 is
type III and 2 are type IV. The 9 southern jet field lines are made
up of 1 type III field line and 8 type IV field lines. The northern
jet is basically a type I event and the southern jet is primarily a
type IV event. Comparing the bottom frame of figure 3 with movie 4
and figure 14, one can understand the white arrows (the azimuthally
averaged poloidal Poynting vector) at the base of the northern jet.
A jet of predominantly type II field lines emerges vertically from
the inner edge of the accretion flow. Once the coiled field lines
reach a vertical elevation above the accretion flow they form a
conical outflow that creates the strong channel of radial Poynting
flux in the bottom frame of figure 3 (the bright red patch) that
seems suspended outside the black hole and above the accretion disk.
The southern jet (that is depicted by the white arrows that emerge
latitudinally from the accretion flow in the bottom frame of figure
3) coincide with a new feature, the white type IV field lines that
appear in figure 14.

\subsection{Overview of Findings}In this section we studied the
distribution of vertical magnetic flux through the equatorial plane
deep in the ergosphere of a rapidly rotating black hole in the
simulation, KDJ. We determined some
potentially interesting results.
\begin{enumerate}
\item We found four distinct topological configurations for vertical
flux through the equatorial plane in the inner ergosphere. The type
1 field lines (figure 6), type II field lines (figure 7), the type
IV field lines (figure 9) are topologies for field lines that
apparently were never envisioned based on theoretical work. The type
III field line configuration (figure 8) was anticipated
theoretically, but it is actually the rarest type of vertical field
line that was found in the KDJ data.
\item All four forms of vertical flux through the inner
accretion disk are associated with the high accretion rate
analog of an ergospheric disk jet \citet{pun07,pun01}.
\item Regions of strong vertical field strength can temporarily
suspend plasma outside of the event horizon. This magnetically
arrested condition is illustrated in figure 12.
\item Bundles of strong magnetic field tends to create low density
regions within the plasma (for type I field lines see animation 2,
for type II field lines see figures 4, 5, 7 and 13).
\item Bundles of strong magnetic field tends to divert the accretion flow towards the opposite hemisphere
(see figures 5, 12 and 13).
\item The vertical magnetic flux distribution is episodic and can
not be described by the time stationary approximation (compare
figures 4, 13 and 14).
\item The vertical magnetic flux and density distribution is inhomogeneous and can
not be described within the context of 2-D.
\end{enumerate}
\par We note that the discovery of vertical flux near the equatorial plane
of the ergosphere in the simulation KDJ is in not accord with the
results presented by one us earlier for a different simulation
\citep{hir04}. Those results were based on 4 different numerical
simulations with spin a/M = 0, a/M = 0.5, a/M = 0.9 and a/M = 0.998.
The simulations in \citet{hir04} were run before KDJ, which is part
of a second generation of simulations in this same family. The only
field line plots in \citet{hir04} that show field lines that don't
penetrate the inner calculational boundary are for KDP, a/M = 0.9.
It is concluded based on a classification scheme of \citet{bla02}
that there are no class 7 field lines, lines that connect the plunge
region to infinity. The type I - IV field lines that we find would
be considered type 7 in that classification scheme. Both field line
integration techniques are based on a fourth order Runge - Kutta
scheme, so it is not the numerical method that accounts for the
difference. Our field lines are not an artifact of the step size in
the Runge - Kutta integration. As we pointed out in section 4.2.1,
variations in the step size showed only minor differences in the
field line integrations, as long as the step size resolution was
comparable to the grid resolution or finer. An occasional field line
might integrate to a different topology with a variation in the
integration step size, but the general situation is unchanged. For
example, the following always hold as long a reasonable (not much
coarser than the grid) step size is chosen, a strong patch of type I
field lines always exists at the same azimuth at t = 9840 M as
indicated in figure 5 and the strong magnetic wall of type II field
lines at t = 9840 M always exists at the same azimuth as indicated
in figure 12. Thus, we conclude that our results differ from
\citet{hir04} because KDP does not have significant vertical flux in
the inner accretion flow and KDJ does. We suspect that the only high
spin simulation considered in \citet{hir04}, KDE (a/M =0.998), might
show some vertical flux in the ergosphere since there is a strong
ergospheric disk jet in this simulation \citep{pun01}.

\par One might find the robustness of these topological identifications surprising in light of
the fact that the disk is a turbulent medium. But, recall that the
field lines in question tend to form magnetic islands in the
accretion flow in regions where the density is typically only $\sim
10\%$ of that in the center of the accretion flow (see sections
4.2.2 and 4.2.3). In these regions of equipartition and stronger
poloidal fields, the MRI turbulence is suppressed \citep{sto01,bal91}. In azimuthally
averaged plots, the ergospheric disk jet is a well ordered region of
strong poloidal field that penetrates the less dense regions of the
accretion flow. The ergospheric jet field stands in obvious
distinction to the adjacent, weaker random patches of turbulent
poloidal field in the main body of the accretion flow. Therefore,
the field line tangling discussed in \citet{hir04} is insignificant
for the field lines described in this paper, except for the portion
of the type I field lines that is restricted to hemisphere that is
opposite of the hemisphere of the jet to which the field line is
linked (see figure 6). More specifically, on the one hand, the
extension of the type I field line toward the jet is not strongly
tangled since it is in a region of strong magnetic field and low
density (see the discussion at the end of section 4.2.3) and
turbulence is minimized. On the other hand, as discussed in section
4.2.2, the type I field lines are anchored to the accretion flow in
the opposite hemisphere and appear as a tangle of azimuthal twists
with themselves and with other azimuthally twisting field lines.
This anchoring end of the type I field lines have tangling
characteristics similar to those found in figure 6 of \citet{hir04}.
However, this is unlikely to affect the jet launching physics to a
significant degree since it lies below the source for the toroidal
field in the jet (i.e., the point were the toroidal magnetic field
changes sign on the type I field line within the disk - where the
Poynting flux and toroidal field in the jet "originates"). The fate
of the endlessly twisted portion of the type I field line does not
affect its topological classification. The tangled end can go into
the horizon, back out into the disk, or out into the corona; it is
still the same topological classification. The tangling of this end
is irrelevant to the topological classification scheme. The only
source of potential misidentification is if there are two type I
field lines adjacent to each other in the accretion flow, one is
coupled to a northern hemisphere jet and one is coupled to a
southern hemisphere jet. Due to the finite grid size and turbulent
mixing, one might misidentify this as a single type III field line.
The fact that an individual time slice rarely shows type I field
lines directed into two different hemispheres implies that this
chance occurrence is rare. Furthermore, the fact that the type III
field lines exit the disk in less than one orbit about the black
hole means that there is minimal opportunity for tangling (see
figure 8). Thus, we do not expect this misidentification to be the
general case, but might occur occasionally if the field type III
field line looked irregular (i.e., large gradients in the tangent
direction) within the accretion flow.

\par In summary, the
topological classification scheme is robust and the small scale
turbulence on the dimension of the grid size does not affect the
classification of type II and IV field lines. It could potentially
lead to the misinterpretation of two oppositely directed type I
field lines that are adjacent within the accretion flow as a single
type III field line. If this rare coincidence happened, it is
probably more of a pedantic distinction that a physical one.
\par With
only three time snapshots, we might not have seen every possible
topological form of vertical flux tube through the inner accretion
flow. It is hard to imagine any other possible topology. However,
that is one of the motivations for performing detailed numerical
work. KDJ has already surprised us by finding type I, type II and
type IV field lines that no one had ever envisioned. It would be
interesting to rerun KDJ with the exact same code and initial state,
but with much higher resolution to see if the higher resolution
induces topological changes (do new types appear? do certain types
disappear?). Also short time interval data dumps would be invaluable
for seeing how the poloidal field evolves. Our findings are
important not for fully categorizing the frequency of all the
possible topological varieties in 3-D accretion flows, but for
opening our eyes to the possible vertical field structures near
accreting black holes within the context of the formation of strong
Poynting jets.

\section{Summary and Discussion}
The dynamic nature of the inner regions of the
accretion flow in KDJ and the simulation in \citet{igu08} are quite
different. Farther out in the main body of the accretion disk MRI driven
turbulent viscosity transports angular momentum outward in both
simulations. However, in \citet{igu08}, the turbulent region is restricted to a range of radii
between the inner disk that is threaded by strong magnetic islands and $r \approx 100 M$ (an azimuthal cross section
of the turbulent region looks similar to the turbulent region in figure 1). A major distinction from
KDJ is that within this turbulent region, the angular momentum is transported outward by both small
scale magnetic torques (MRI turbulence) and large scale magnetic torques (associated with
the net vertical flux through the equatorial plane)
and in principle it is difficult to distinguish between the two.
In the inner regions, it is large scale magnetic torques
associated with the magnetic islands that is primarily responsible for
extracting angular momentum from the plasma in \citet{igu08}. In
KDJ, for most of the inflowing plasma near the black hole, turbulent viscosity
torques the ingoing plasma except in localized magnetic islands where the ergospheric disk jet is launched. Near
and within these magnetic islands, the dynamics of angular
momentum redistribution is similar to \citet{igu08}. In spite of these
differences, we found many results that are common to both extremes
in black hole spin rate, as can be seen in the list of results,
below.

\begin{enumerate}

\item All four types of vertical field lines around a rapidly rotating black hole support strong outward
Poynting flux. There is also modest amounts of Poynting flux that is
driven from the accretion flow along the vertical field lines in the
Pseudo-Newtonian potential (see figure 3 for the Kerr case and
\citet{igu08} for the Pseudo-Newtonian potential).

\item In regions where vertical flux builds up in the inner accretion flow, the mass density goes down within that same region
(see figure 2 for the Pseudo-Newtonian potential and figures 4, 5,
12 and 13 for the Kerr case).

\item The vertical flux build up (the formation of magnetic islands) is episodic and
inhomogeneous. The distribution of vertical magnetic field in 3-D
can not be described in the time stationary approximation (see
figure 2 for the Pseudo-Newtonian potential and figures 4, 5, 11, 13
and 14 for the Kerr case).

\item The vertical flux distribution is
inhomogeneous. The distribution of vertical magnetic field in 3-D
can not be described in the 2-D approximation (see figure 2 for the
Pseudo-Newtonian potential and figures 4, 5, 11, 13 and 14 for the
Kerr case).

\item The flow can be temporarily arrested (dr/dt=0) within and adjacent to the region of vertical flux build up (see
figure 2 and \citet{igu08} for the Pseudo-Newtonian potential and
figure 12 for the Kerr case).

\item The flow is diverted by the magnetic pressure of the magnetic islands and tends to accrete in narrow channels
or "spiral streams" that flow around the magnetic islands (see
figures 2 for the Pseudo-Newtonian potential and figures 4, 5, 12,
13 and the comment on the comparison of figures 4 and 5 with figure
8 in section 4.2.2 for the Kerr case).

\end{enumerate}
The reason for the commonality within the list is clear. The
existence of magnetic islands near a black hole is not a consequence
of the dragging of inertial frames or the detailed history of how
the magnetic flux reached the black hole to first order. MRI
turbulence does not seem to produce significant Poynting jets in
accretion flows as indicated in the test case simulations in
\citet{igu03,bec08}. It is the advection of a net poloidal field
that creates the magnetic islands and the associated Poynting jet.
In \citet{igu08} this last statement is obvious by construction. In
the poloidal loop simulations this is a bit more obscure. The field
in the funnel is created by the leading edge of the accreted
poloidal loops at the beginning of the simulation. The poloidal
field orientation in the funnel at late times is always the same as
the direction of the leading edge of the poloidal loops in the
initial state torus (K. Beckwith and J. Hawley private communication
2008). The flux is trapped within the vortex and over time there are
random fluctuations in the field distributions due to the MRI
created poloidal fields that are considerably smaller, but the net
radial field does not change sign. The poloidal field in the funnel
is much stronger than the MRI generated poloidal field in the
accretion flow. In KDJ, the strongest poloidal field is always at
the base of the ergospheric disk jet and in the same direction as
the funnel field. Thus, in KDJ it appears that it is the initial
seed poloidal loops that were accreted that are ultimately the
source of the poloidal flux in the magnetic islands in the inner
accretion flow. This is consistent with results of the initial state
parametric study performed in \citet{bec08}. If the leading edge of
the initial state poloidal loops is destroyed by reconnection in the
funnel or not there by construction, the funnel field and jet become
insignificant or nonexistent \citep{bec08}. The commonality between
KDJ and the simulations in \citet{igu08} listed above arise because
both simulations have a net vertical flux accreted from distant
regions into the immediate vicinity of the black hole.

\par The primary difference between the Pseudo-Newtonian case and the
Kerr example is that the field is much more twisted due to strong
frame dragging forces in the Kerr geometry. This has two primary
effects. First, the azimuthal twists make the field lines look much
more intricate and the resulting complexity of the magnetosphere is
much greater. Another major difference is the amount of Poynting
energy that is transported by the vertical field, there is at least
an order of magnitude more in the Kerr case (about 1\% - 2\% of the
accretion flow binding energy is transformed into Poynting flux in
the Pseudo-Newtonian potential, \citet{igu08}, and about 25\% of the
accretion flow binding energy is transformed into Poynting flux in
KDJ, \citet{haw06,pun07}).
\par There is another large difference
that is not dependent on the black hole spin, but the assumptions of
the simulations, there is no net vertical flux in KDJ and in the
Pseudo-Newtonian simulations there is an infinite reservoir of
vertical flux that is accreted. The end result is that there is much
more vertical flux near the black hole in the Pseudo-Newtonian
simulation. Presumably, accreting vertical flux from the outer
boundary in the $a/M=0.99$ Kerr spacetime background would also
result in an enhancement of vertical flux near the black hole
relative to what was found in KDJ. We believe that if this were
true, the vertical flux distribution and the accretion dynamics in
the high spin case would closely resemble the simulations of
\citet{igu08}. Presumably such an enhancement of vertical flux near
the black hole would increase the extent of the ergospheric disk and
the degree of magnetization. However, because of the nonlinear
evolution of such a numerical system, a 3-D simulation of the
$a/M=0.99$ black hole with vertical flux accretion from the outer
boundary needs to be run to see if there is merit to these
speculations. Figure 11.16 of \citet{pun01} is a plot based on the
KDJ data that indicates that the jet power from the ergospheric disk
is potentially 5 - 50 times more than can be derived from the
Blandford-Znajek process that is associated with the event horizon
magnetosphere. This ratio depends on spin rate and the degree to
which the ergospheric disk is saturated with vertical flux. For
example, with the degree of vertical flux saturation in KDJ with
a/M=0.99, results in a ratio of ergospheric disk to Blandford-Znajek
power of $\gtrsim 5 $, for a completely saturated ergospheric disk
with a/M = 0.995, the ratio is $\approx 50$. This result is also
apparent from figure 3, where most of the power emerges in the
narrow, red/yellow channels that are fed by the ergospheric disk.
Without the ergospheric disk, one just has the blue background (the
Blandford-Znajek power for the event horizon) in figure 3. An
extremely efficient powerful jet seems to be a requirement of many
powerful FRII radio sources \citep{pun10}. In particular, vertical
flux through the equatorial plane seems to be a viable candidate to
support the more powerful FRII jets from low accretion rate
(kinetically dominated) systems (see also \citet{nem07} for similar
conclusions).
\par In summary, in this paper we explored vertical magnetic flux
in simulations of accretion flows near black holes. We categorized
the magnetic field line topologies that were found and discussed
conditions within the simulations that give rise to magnetic islands
near black holes. One might wonder if there is any physical
significance to the existence of vertical flux near a black hole or
is it merely a pedantic exercise to distinguish between vertical
flux and an axisymmetric event horizon magnetosphere. The motivation
to differentiate the vertical flux from the horizon flux is
two-fold. First of all, the strong interaction between vertical flux
and the accretion flow has no counter-part in an event horizon
magnetosphere located within the accretion vortex. The strong
interaction was shown in this paper and in \citet{igu08} to lead to
episodic, inhomogeneous field configurations. It was posited in
\citet{igu08} that this could be related to quasi-periodic
oscillations in black hole binaries. Furthermore, the non-time
stationary behavior could be related to high energy flares in AGN.
Secondly, the existence of vertical flux increases the power of the
Poynting jets from black hole accretion systems.

\par As a final comment, we mention the interesting 3-D numerical
simulations performed by \citet{fra07}. They have modeled accretion
flows that are slightly inclined with respect to the equatorial
plane. Surprisingly, it was found that merely tilting the accretion
flow by $15^{\circ}$ from the equatorial plane that the majority of
the gas does not arrive at the equator of the event horizon, but at
high latitudes near the poles. This creates low density cavities in
the equatorial plane of the ergosphere. Now consider figures 5, 12 and
13 that seem to indicate that in KDJ, the existence of vertical
flux that emerges from an accretion flow is correlated with the
accretion flow being displaced toward the opposite hemisphere and
the formation of low density cavities near the equatorial plane.
Thus, it would be interesting to see if future higher resolution
simulations of the tilted accretion disk naturally develop
significant vertical flux emerging from the inner accretion flow
within the ergosphere. However, we caution the reader that this code
is quite different from that used in KDJ. Because of the lack of
symmetry in the problem, it is difficult to get high numerical
resolution in all the dynamically important regions. The simulations
are likely to be too numerically diffusive for our goals and the
results might not be a reliable indicator of magnetic flux
evolution. A more reliable understanding of magnetic flux evolution
in the tilted disk case might have to await future generations of
simulations.
\appendix
\section{Appendix. Field Line Integration Method} We describe our
method of plotting field lines in 3-D using Paraview 3.3.0. This is
nontrivial since the Boyer-Lindquist coordinate system is tantamount
to a differentiable mapping of the spherical Minkowski coordinates,
$(t,\; r,\; \theta ,\; \phi)$ defined in $R^{4}$ into the curved
spacetime outside of inner calculational boundary of the
computational grid. In order to display three dimensional geometries
(such as the magnetic field fibration of a subset of the spacetime
outside the inner calculational boundary) embedded in the curved
Kerr spacetime requires imaging in the our flat space 3-D world,
$R^{3}$. Thus, this mapping into $R^{3}$ for visualization purposes
is not an isometric mapping from the Kerr spacetime. In particular,
the 3-D visualization software Paraview only reads the magnetic
field data in flat space Cartesian coordinates,
\begin{equation}
\mathbf{B}= (B^{x}(x,y,z),\; B^{y}(x,y,z),\; B^{z}(x,y,z))\;.
\end{equation}
Paraview uses a Runge-Kutta scheme to integrate the following
ordinary differential equation to draw field lines $(x(s), y(s),
z(s))$:
\begin{eqnarray}
&& \frac{dx}{ds} = \frac{B^{x}}{\|B\|} \\
&& \frac{dy}{ds} = \frac{B^{y}}{\|B\|} \\
&& \frac{dz}{ds} = \frac{B^{z}}{\|B\|},
\end{eqnarray}
where
\begin{eqnarray}
&& \|B\|^{2} \equiv (B^{x})^{2} + (B^{y})^{2} + (B^{z})^{2}\\
&& ds^{2} \equiv dx^{2} + dy^{2} + dz^{2}.
\end{eqnarray}
On the other hand, the Boyer-Lindquist coordinate system is
analogous to a spherical coordinate system in which the magnetic
field components are given in terms of the Faraday tensor in
\citet{pun01} as
\begin{eqnarray}
&& \mathbf{B}= (B^{r}(r,\; \theta,\; \phi),\; B^{\theta}(r,\;
\theta,\; \phi),\; B^{\phi}((r,\; \theta,\; \phi))= \nonumber\\
&& (F_{\theta \phi}(r,\; \theta,\; \phi),\;  F_{\phi r}(r,\;
\theta,\; \phi),\; F_{r \theta}(r,\; \theta,\; \phi))\;.
\end{eqnarray}
In Boyer-Lindquist coordinates, the magnetic field lines are defined
by the following system of equations
\begin{eqnarray}
&& \frac{dr}{ds} = \frac{B^{r}}{\|B\|} \\
&& \frac{d\theta}{ds} = \frac{B^{\theta}}{\|B\|} \\
&& \frac{d\phi}{ds} = \frac{B^{\phi}}{\|B\|}\;,
\end{eqnarray}
where
\begin{eqnarray}
&& \|B\|^{2} \equiv g_{rr}{B^{r}}^{2} + g_{\theta\theta}{B^{\theta}}^{2} + g_{\phi\phi}{B^{\phi}}^{2}\\
&& ds^{2} \equiv g_{rr}dr^{2} + g_{\theta\theta}d\theta^{2} +
g_{\phi\phi}d\phi^{2}.
\end{eqnarray}
One can realize the Boyer-Lindquist magnetic field line definition
in equations (A.8) - (A.12) in the context of flat space Cartesian
coordinates (suitable for Paraview 3.3.0) by differentiating the
transformation from Cartesian to spherical coordinates,
\begin{eqnarray}
&& x \equiv r\sin\theta\cos\phi \\
&& y \equiv r\sin\theta\sin\phi \\
&& z \equiv r\cos\theta \;,
\end{eqnarray}
on both sides by the affine parameter "ds". Then the quantities,
$dr/ds$, $d\theta/ds$ and $d\phi/ds$, that appear after
differentiation by "ds" can be replaced with the right hand side of
(A.8) - (A.10) to get
\begin{eqnarray}
&& \frac{dx}{ds} = \frac{(B^{r})\sin\theta\cos\phi + (rB^{\theta})\cos\theta\cos\phi - (r\sin\theta B^{\phi})\sin\phi}{\|B\|} \\
&& \frac{dy}{ds} = \frac{(B^{r})\sin\theta\sin\phi + (rB^{\theta})\cos\theta\sin\phi + (r\sin\theta B^{\phi})\cos\phi}{\|B\|} \\
&& \frac{dz}{ds} = \frac{(B^{r})\cos\theta -
(rB^{\theta})\sin\theta}{\|B\|}\;.
\end{eqnarray}
We integrate equations (A.16) - (A.18) with Paraview 3.3.0 for the
field line data from the simulation KDJ.

\begin{acknowledgements}
We would like to thank Jean-Pierre DeVilliers for sharing his deep
understanding of the numerical code and these simulations. We were
also very fortunate that Julian Krolik and John Hawley were willing
to share their data in the best spirit of science. This study was
partially supported by the U.S. Department of Energy (DOE) Office of
Inertial Confinement Fusion under Cooperative Agreement No.
DE-FC52-08NA28302, the University of Rochester, the New York State
Energy Research and Development Authority.
\end{acknowledgements}


\begin{thebibliography}{}
\bibitem[Archontis and Hood(2008)]{arc08}Archontis, V., Hood, A., 2008, ApJ \textbf{647} 113
\bibitem[Balbus and Hawley(1991)]{bal91}Balbus, S. A., \& Hawley, J. F. 1991, ApJ \textbf{376} 214
\bibitem[Beckwith et al(2008)]{bec08}Beckwith, K., Hawley, J., Krolik, J. 2008, ApJ \textbf{678} 1180
\bibitem[Bisnovatyi-Kogan and Ruzmaikin(1974)]{bis74}Bisnovatyi-Kogan, G. S. and Ruzmaikin, A. A. 1974, Ap \& SS \textbf{28} 45
\bibitem[Bisnovatyi-Kogan and Ruzmaikin(1976)]{bis76} Bisnovatyi-Kogan, G. S. and Ruzmaikin, A. A. 1976, Ap \& SS \textbf{42} 401
\bibitem[Blandford(1976)]{bla76}Blandford, R. D. 1976, MNRAS \textbf{176} 465
\bibitem[Blandford(2002)]{bla02}Blandford, R. D. 2002, in \textbf{Lighthouses of the Universe: The Most Luminous
Celestial Objects and Their Use for Cosmology}, ed. M. Gilfanov, R.
Sunyaev, \& E. Churazov (New York: Springer), 381
\bibitem[Blandford and Payne(1982)]{bla82}Blandford, R. D. and Payne,D., 1982, MNRAS \textbf{199}
883
\bibitem[Blandford and Znajek(1977)]{blz77} Blandford, R. and Znajek,
R. 1977, MNRAS. \textbf{179}, 433
\bibitem[Cabral and Leedom(1993)]{cab93} Cabral, B., \& Leedom, L. 1993, Computer Graphics: Proceedings:
  Annual Conference Series 1993: SIGGRAPH 93 (New York: Association for
  Computing Machinery), 263
\bibitem[Colella and Woodward (1984)]{col84} Colella, P., \& Woodward, P.R. 1984, J. Comp. Phys., 54, 174
\bibitem[De Villiers and Hawley (2003a)]{dev02} De Villiers, J., Hawley, 2003,
 ApJ \textbf{589}, 458
\bibitem[De Villiers et al(2003b)]{dev03} De Villiers, J-P., Hawley, J., Krolik, 2003,
 ApJ \textbf{599} 1238
\bibitem[De Villiers et al(2005)]{dev05} De Villiers, J-P., Hawley, J., Krolik, J.,Hirose, S.
2005, ApJ \textbf{620} 878
\bibitem[Dorch  et al(1999)]{dor99}Dorch, S., Archontis, V., Nordlund, A.,1999, A \& A \textbf{352} L79
\bibitem[Fragile et al(2007)]{fra07}Fragile, P. C., Blaes, O. M., Anninos, P., Salmonson, J. D. 2007, ApJ \textbf{668} 417
\bibitem[Gardiner and Stone(2005)]{gar05} Gardiner, T.A., \& Stone, J.M. 2005, J. Comp. Phys., 205, 509
\bibitem[Ghosh and Abramowicz(1997)]{pra97} Ghosh, P., \& Abramowicz, M. A. 1997, MNRAS \textbf{292}, 887
\bibitem[Hawley and Krolik (2006)]{haw06} Hawley, J., Krolik, K.
2006, ApJ \textbf{641} 103
\bibitem[Hirose et al (2004)]{hir04}Hirose, S., Krolik, K., De
  Villiers, J., Hawley, J. 2004, ApJ \textbf{606}, 1083
\bibitem[Igumenshchev(2008)]{igu08}Igumenshchev, I. V. 2008, ApJ \textbf{677} 317
\bibitem[Igumenshchev et al(2003)]{igu03}Igumenshchev, I. V., Narayan, R. and Abramowicz, M. A. 2003, ApJ \textbf{592} 1042
\bibitem[Kantrowitz and Petschek(1966)]{kap66} Kantrowicz, A.R. and
  Petschek, H.E. 1966, In: \emph{Plasma Physics in Theory and Application.} ed. by W.B. Kunkel (McGraw-Hill, New York) p. 148
\bibitem[Kato et al(2004)]{kat04}Kato, Y., Mineshige, S. and Shibata, K. 2004, ApJ \textbf{605} 307
\bibitem[Komissarov(2004)]{kom04}Komissarov, S. 2004, MNRAS \textbf{350} 1431
\bibitem[Komissarov and McKinney(2007)]{kom07}Komissarov, S., McKinney, J. 2007, MNRAS \textbf{377} 49
\bibitem[Krolik et al (2005)]{kro05} Krolik, K., Hawley, J., Hirose, S.
2005, ApJ \textbf{622}, 1008
\bibitem[Kulkarni and Romanova (2008)]{kul08}Kulkarni, A., Romanova, M. M. 2008, MNRAS \textbf{386} 673
\bibitem[Landau and Lifshitz (1987)]{lan87} Landau, L. D., \& Lifshitz, E. M. 1987, Electrodynamics of
  Continuous Media. Pergamon Press, Oxford
\bibitem[Lovelace(1976)]{lov76}Lovelace, R. V. E. 1976, Nature \textbf{262} 649
\bibitem[Lubow et al(1994)]{lub94}Lubow, S. H., Papaloizou, J. C. B and  Pringle, J. E. 1994, MNRAS \textbf{267} 235
\bibitem[McKinney and Gammie(2004)]{mck04}McKinney, J. and Gammie, C. 2004, ApJ \textbf{611} 977
\bibitem[McKinney(2005)]{mck05}McKinney, J. 2004, ApJL\textbf{630} 5
\bibitem[Meier(2004)]{mei04}Meier, D. L. 2004, ApJ \textbf{605} 340
\bibitem[Narayan et al(2003)]{nar03}Narayan, R., Igumenshchev, I. V. and Abramowicz, M. A. 2003, PASJ \textbf{55} L69
\bibitem[Nemmen et al(2007)]{nem07}Nemmen, R.,Bower, R.,  Babul, A., Storchi-Bergmann, T. 2007, MNRAS \textbf{377}
1652
\bibitem[Paczy\'nski and Wiita (1980)]{pac80} Paczy\'nski, B., \& Wiita, P.J. 1980, A\&A, 88, 23
\bibitem[Punsly and Coroniti(1990)]{pun90}Punsly, B., Coroniti, F.V. 1990, ApJ \textbf{354} 583
\bibitem[Punsly(1996)]{pun96}Punsly, B. 1996, ApJ \textbf{467} 105
\bibitem[Punsly(2008)]{pun01} Punsly, B. 2008, \emph{Black Hole
Gravitohydromagnetics}, second edition (Springer-Verlag, New York)
\bibitem[Punsly(2007a)]{pun07}Punsly, B. 2007, ApJL \textbf{661}, 21
\bibitem[Punsly(2007b)]{pun08}Punsly, B. 2007, MNRAS Letters \textbf{381},79
\bibitem[Punsly(2007c)]{pun10}Punsly, B. 2007, MNRAS \textbf{374} 10
\bibitem[Reynolds et al (2006)]{rey06}Reynolds,C., Garofalo, D.and Begelman, M. 2006, ApJ \textbf{651} 1023
\bibitem[Romanova et al (2008)]{rom08}Romanova, M. M., Kulkarni, A. and Lovelace, R. V. E. 2008, ApJ \textbf{673} 171
\bibitem[Rothstein and Lovelace(2008)]{rot08}Rothstein, D. and Lovelace, R. V. E. 2008, ApJ \textbf{677} 1221
\bibitem[Semenov et al (2004)]{sem04} Semenov, V., Dyadechkin, S. and Punsly, B. 2004,
Science \textbf{305}978
\bibitem[Shibata et al(1990)]{shi90}Shibata, K., Tajima, T., Matsumoto, R. 1990, ApJ \textbf{350} 295
\bibitem[Spruit and Uzdensky (2005)]{spr05}Spruit, H.C. and
Uzdensky, D.A. 2005, ApJ \textbf{629} 960
\bibitem[Stone and Pringle (2001)]{sto01}Stone, J. M. and Pringle, J. 2001, MNRAS, \textbf{322}, 461
\bibitem[Thorne et al(1986)]{thp86} Thorne, K., Price, R. and Macdonald, D. 1986,
\emph{Black Holes: The Membrane Paradigm} (Yale University Press,
New Haven)
\bibitem[Ustyugova et al (2000)]{ust00}Ustyugova, G. V., Lovelace, R. V. E., Romanova, M. M., Li, H and Colgate, S. 2000, ApJL \textbf{541} 21
\bibitem[van Ballegooijen(1989)]{van89}van Ballegooijen, A. A. 1989, in ASSL Vol. 156 \textbf{Accretion Disks and Magnetic Fields in Astrophysics},
ed. G. Belvedere (Dordrecht: Kluwer Academic Publishers), 99
\bibitem[Warren et al (2008)]{war08} Warren, H. et al, 2008, ApL \textbf{686}, 131
\end{thebibliography}
\end{document}